%% file: main.tex
\documentclass[]{research}
\input{math_commands.tex}

\usepackage{hyperref}
\usepackage{url}
\usepackage{adjustbox} 
\usepackage{multirow}
\usepackage{booktabs} 
\usepackage{pifont}
\usepackage{subcaption} 
\usepackage{makecell}
\usepackage{wrapfig}
\usepackage{float}
\usepackage{calc}
\usepackage{enumitem}
\usepackage{xcolor}
\usepackage{colortbl}
\usepackage{listings}
\usepackage{tcolorbox}
\usepackage{amsmath,amssymb}
\usepackage{algorithm}
\usepackage{algpseudocode}

\title{LunaTP: Enabling Efficient and Scalable Tensor Product for Equivariant MLIPs}

\author[1,2]{Yuanchang Zhou}
\author[1,2,*]{Mingzhen Li}
\author[1,2]{Chen Wang}
\author[1,2]{Xiangyu Zhang}
\author[1,2]{Guangming Tan}
\author[1,2,*]{Weile Jia}

\affiliation[1]{State Key Lab of Processors, Institute of Computing Technology, Chinese Academy of Sciences}
\affiliation[2]{University of Chinese Academy of Sciences}
\contribution[*]{Corresponding author}
\metadata[Email]{%
\href{mailto:zhouyuanchang23s@ict.ac.cn}{zhouyuanchang23s@ict.ac.cn},
\href{mailto:limingzhen@ict.ac.cn}{limingzhen@ict.ac.cn},
\href{mailto:jiaweile@ict.ac.cn}{jiaweile@ict.ac.cn}}
\metadata[Code]{\url{https://github.com/HPC-AI-Team/LunaEquivariance}}

\abstract{
Equivariant machine learning interatomic potentials (MLIPs) promise atomistic simulations with accuracy approaching that of quantum-mechanical calculations. However, the high computational cost of their core Clebsch--Gordan tensor product (CGTP) convolutions severely limits model scalability and efficiency. In this paper, we rethink the parallel execution of CGTP convolutions and introduce LunaTP. Built on node-centric execution, LunaTP exploits path-group parallelism and multi-level reuse to accelerate CGTP convolutions and higher-order derivative computation while preserving the underlying mathematical formulation. On an NVIDIA H100 GPU, LunaTP achieves kernel speedups of up to 250.0$\times$ over e3nn and 17.1$\times$ over OpenEquivariance. It also delivers end-to-end training throughput gains of up to 28.7$\times$ over e3nn and 3.11$\times$ over OpenEquivariance.
Finally, LunaTP demonstrates scalability and robust performance across models, workload scales, and GPUs, delivering substantial acceleration for training and atomistic simulation with high-accuracy equivariant MLIPs.
}

\begin{document}

\maketitle

\section{Introduction}\label{sec: intro}
Scaling equivariant machine learning interatomic potentials (MLIPs) is challenging, since the Clebsch-Gordan tensor products (CGTP) at their heart have computational cost that scales as $O(L_{max}^6)$ in the maximum degree $L_{max}$ \citep{3618408.3619548, luo2024enabling}.
Consequently, modern equivariant MLIPs are often restricted to \(L_{\max}\) to small values (typically no greater than 3), limiting the use of high-degree equivariant features \citep{batatia2022mace, liao2023equiformer, Park_2024}. Ideally, we would like to use higher-degree features to better capture angular resolution and directional information, which are crucial for accurate potential energy surface (PES) modeling \citep{Batzner_2022, liao2024equiformerv2improvedequivarianttransformer, duval2024hitchhikersguidegeometricgnns, xie2025the}.
\begin{wrapfigure}[17]{r}{0.5\textwidth}
    \centering
    \includegraphics[width=0.92\linewidth]{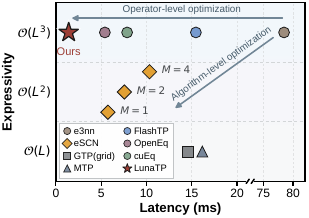}
    \caption{
    Expressivity--efficiency landscape of tensor product convolutions. Forward latency is measured on an NVIDIA H100 with 768 atoms, $L_{\max}=6$, and 64 channels per degree.
    }
    \label{fig: intro expressivity}
\end{wrapfigure}

In modern equivariant MLIPs \citep{Batzner_2022, kim_optimizing_2026, xu2026spectralspatialtensoratomiccluster,kavanagh2026fastaccuratefoundationmodels}, CGTP convolutions can account for over 70\% of the training time, making them a computational bottleneck for model scaling. A series of pioneering works have explored ways to optimize CGTP. Their approaches fall into two categories (see Figure \ref{fig: intro expressivity}). Algorithm-level methods reformulate the tensor product or reduce its degrees of freedom, resulting in a trade-off between expressivity and efficiency \citep{unke2024e3xmathrme3equivariantdeeplearning, luo2024enabling, 3618408.3619548, xie2025the}. By contrast, operator-level methods improve the efficiency of CGTP on GPUs without changing its mathematical formulation or sacrificing expressivity through kernel optimizations \citep{nvidia_cuequivariance, lee2025flashtp, bharadwaj2025efficientsparsekernelgenerator}. Our work follows this direction and introduces LunaTP to further improve the efficiency of CGTP convolutions.

However, further accelerating CGTP convolutions while preserving the full CGTP formulation remains challenging. Although kernel fusion and sparse CG computation reduce computational cost and memory traffic~\citep{nvidia_cuequivariance, lee2025flashtp, bharadwaj2025efficientsparsekernelgenerator}, edge-parallel execution still suffers from atomic write contention when multiple edges accumulate messages at the same destination node. Node-centric reduction can eliminate this contention, but does not necessarily improve overall performance. Variations in node degree and per-path computational cost complicate load balancing, while increasing data reuse across paths raises on-chip resource requirements and can limit parallelism. Further improving CGTP convolution performance therefore requires jointly optimizing reduction, data reuse, and parallel execution.

To bridge this performance gap, we introduce LunaTP, which improves the execution of CGTP convolutions without sacrificing expressivity. Our main contributions are:
\begin{itemize}
    \item We rethink the parallel execution of CGTP convolutions and introduce a node-centric execution scheme that accumulates edge messages on chip, reducing global memory updates.

    \item Building on this execution scheme, we develop path-group parallelism to balance input reuse and register usage, and reduce redundant computation through sparse expansion and multi-level reuse.

    \item We extend this execution scheme to higher-order backward passes, enabling efficient force- and Hessian-supervised training.

    \item We evaluate LunaTP through kernel benchmarks and end-to-end experiments on representative MLIP workloads. On an NVIDIA H100, LunaTP achieves up to a 250.0$\times$ kernel speedup and 28.7$\times$ higher training throughput than e3nn.
\end{itemize}

\section{Preliminaries}\label{sec: preliminary} 
\subsection{Clebsch--Gordan Tensor Product}\label{sec:cgtp}
The Clebsch--Gordan tensor product (CGTP) computes each degree-$l$ component in the direct-sum decomposition.
Given two irreducible representations (irreps)
$\mathbf{x}^{(l_1)}\in\mathbb{R}^{2l_1+1}$ and
$\mathbf{y}^{(l_2)}\in\mathbb{R}^{2l_2+1}$, the $m$-th component of the degree-$l$ output is given by:
\begin{equation}
    \left[
        \mathbf{x}^{(l_1)}
        \otimes
        \mathbf{y}^{(l_2)}
    \right]^{(l)}_m
    =
    \sum_{m_1=-l_1}^{l_1}
    \sum_{m_2=-l_2}^{l_2}
    C^{(l,m)}_{(l_1,m_1)(l_2,m_2)}
    x^{(l_1)}_{m_1}
    y^{(l_2)}_{m_2},
\label{eq:cgtp}
\end{equation}
where $C^{(l,m)}_{(l_1,m_1)(l_2,m_2)}$ are the CG coefficients and are nonzero only when $|l_1-l_2|\leq l\leq |l_1+l_2|$, making the coefficient matrices more than 70\% sparse \citep{bharadwaj2025efficientsparsekernelgenerator,lee2025flashtp}.

We refer to each coupling from input degrees $l_1$ and $l_2$ to an output degree $l$ as a \emph{path}, denoted by $(l_1,l_2)\to l$. Each path can be assigned a separate learnable weight while preserving equivariance. For irreps up to degree $L$, the full tensor product contains $\mathcal{O}(L^3)$ paths. In a naive implementation, each path requires $\mathcal{O}(L^3)$ multiply--add operations, yielding $\mathcal{O}(L^6)$ total complexity.

To understand how per-path CGTPs are combined in equivariant MLIPs, we next describe the tensor product convolution layer.

\subsection{Tensor Product Convolution}
A tensor product convolution computes one CGTP for every combination of edge, channel, and path \citep{geiger2022e3nneuclideanneuralnetworks,Batzner_2022,batatia2022mace,Park_2024}. With $E$ edges, $C$ channels, and $P$ paths, the layer performs a total of $E\times C\times P$ such computations. The resulting edge messages are then aggregated at their destination nodes.

\begin{algorithm}[h]
\caption{Standard Clebsch--Gordan Tensor Product Convolution}
\label{alg:tpconv}
\begin{algorithmic}[1]
\Statex \textbf{Input:} $\mathbf{x}_j^{(l_1,c)}$: node irreps
\Statex \textbf{Input:} $\mathbf{y}_{ji}^{(l_2)}$: angular edge irreps
\Statex \textbf{Input:} $w_{ji}^{(l_1,l_2,l,c)}$: path weights generated by the radial MLP
\Statex \textbf{Input:} edges $ji$, from source node $j$ to destination node $i$
\Statex \textbf{Output:} ${\mathbf{x}'}_i^{(l,c)}$: updated node irreps
\State Initialize ${\mathbf{x}'}_i\leftarrow\mathbf{0}$ for all nodes $i$
\For{$ji$ in the $E$ edges} \Comment{Compute edge messages}
    \State Initialize $\widetilde{\mathbf{m}}_{ji}\leftarrow\mathbf{0}$
    \For{$c$ in the $C$ channels} \Comment{Process each feature channel}
        \For{$(l_1,l_2)\to l$ in the $P$ paths} \Comment{Compute a CGTP for each path}
            \State $\widetilde{\mathbf{m}}_{ji}^{(l,c)}
            \leftarrow
            \widetilde{\mathbf{m}}_{ji}^{(l,c)}
            +
            w_{ji}^{(l_1,l_2,l,c)}
            \left[
                \mathbf{x}_j^{(l_1,c)}
                \otimes
                \mathbf{y}_{ji}^{(l_2)}
            \right]^{(l)}$
        \EndFor
    \EndFor
    \State ${\mathbf{x}'}_i
    \leftarrow
    {\mathbf{x}'}_i+\widetilde{\mathbf{m}}_{ji}$
    \Comment{Aggregate at destination node $i$}
\EndFor
\end{algorithmic}
\end{algorithm}
Algorithm~\ref{alg:tpconv} presents the pseudocode for the tensor product convolution. For each edge $ji$, the CGTP couples the source-node irrep $\mathbf{x}_j^{(l_1,c)}$ with the angular edge irrep $\mathbf{y}_{ji}^{(l_2)}$. The result is a full degree-$l$ output, whose $m$-th component is given by Eq.~\ref{eq:cgtp}. This output is scaled by the corresponding path weight $w_{ji}^{(l_1,l_2,l,c)}$ and accumulated into the edge message (Line~6). The same operation is performed for every channel--path combination to construct the edge message $\widetilde{\mathbf{m}}_{ji}$ (Lines~2--8). The message is then aggregated at destination node $i$ (Line~9). 
In particular, training conservative MLIPs typically requires higher-order derivatives, and we provide the corresponding pseudocode in Appendix~\ref{app: derivation of tp conv}.

While the standard implementation is mathematically straightforward, its performance on modern GPUs depends on how the computation is organized, as we discuss next.

\section{Rethinking Parallelism for CGTP Convolution}\label{sec: motivation}
\begin{figure}[h]
\begin{center}
    \includegraphics[width=1.0\linewidth]{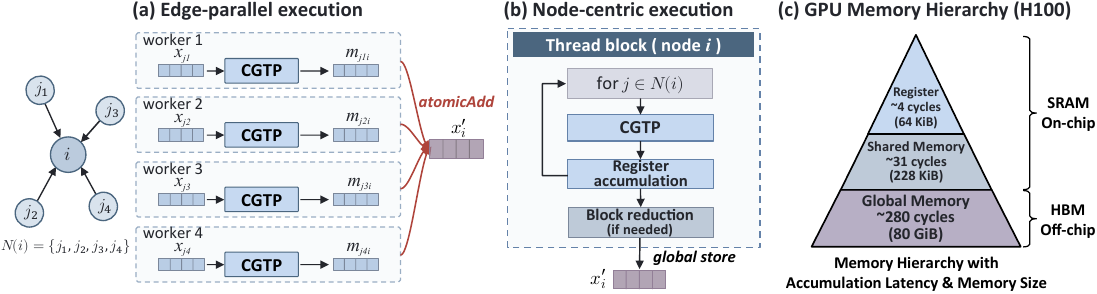}
    \caption{ 
    CGTP convolution parallelism and accumulation latencies across the memory hierarchy.
    }
    \label{fig: motivation}
\end{center}
\end{figure}
Parallelism across edges is a common execution scheme in CGTP convolutions (Figure~\ref{fig: motivation}(a))~\citep{lee2025flashtp, bharadwaj2025efficientsparsekernelgenerator}. Since CGTP on each edge are independent, this scheme exploits data parallelism by assigning per-edge computations to different workers. However, during message aggregation (Algorithm~\ref{alg:tpconv}, line 9), many workers attempt to update the same node concurrently, causing atomic contention and hardware-level serialization. 
Our profiling on an NVIDIA H100 GPU shows that aggregation can account for more than 30\% of the total runtime. Thus, although kernel fusion and sparse CG evaluation reduce intermediate memory accesses and tensor product computation, aggregation still incurs substantial overhead.

On modern GPUs, the latency of accumulation depends on where the data resides. As Figure~\ref{fig: motivation}(c) shows, accumulation in registers and shared memory typically has lower latency than global atomic updates. This raises a natural question: can we use these low-latency on-chip memories to perform neighbor aggregation?

Node-centric execution provides a natural way to do so (Figure~\ref{fig: motivation}(b)). This scheme aggregates messages for each node on chip and writes the aggregated result to global memory. 
A straightforward way to implement this is to assign one thread block to each node. However, our experiments show that this direct mapping eliminates per-edge atomic updates but does not improve runtime, as it does not fully exploit intra-node parallelism. Thus, efficient node-centric execution still faces the following primary challenges:

\noindent\textbf{(1) Work partitioning limited by on-chip reduction.}
Node-centric execution requires each thread block to traverse all edges of a node and aggregate messages on chip. This makes work partitioning difficult, since the workload depends on both node degree and path computation. On the one hand, node degree determines how many edges a block traverses (e.g., degree distributions are often skewed in atomistic datasets~\citep{barros-luque_open_2026,levine2026openmolecules2025omol25,gharakhanyan_open_2026}). On the other hand, the amount of CG computation also varies considerably across paths. Thus, blocks computing the same number of output features may still have different workloads.

\noindent\textbf{(2) Fusion scope limited by register capacity.}
In node-centric execution, we aim to exploit shared inputs across paths to reduce redundant memory accesses and computation. However, each fused path needs to keep its output accumulators live throughout neighbor reduction, so a larger fusion scope increases register demand and may lower occupancy or cause spilling. Under identical thread configuration and register constraints, fusing all compatible paths runs 35.4\% slower than limiting the fusion scope and incurs register spilling.

\section{LunaTP}\label{sec: methods}
\begin{figure}[t]
\begin{center}
    \includegraphics[width=1.0\linewidth]{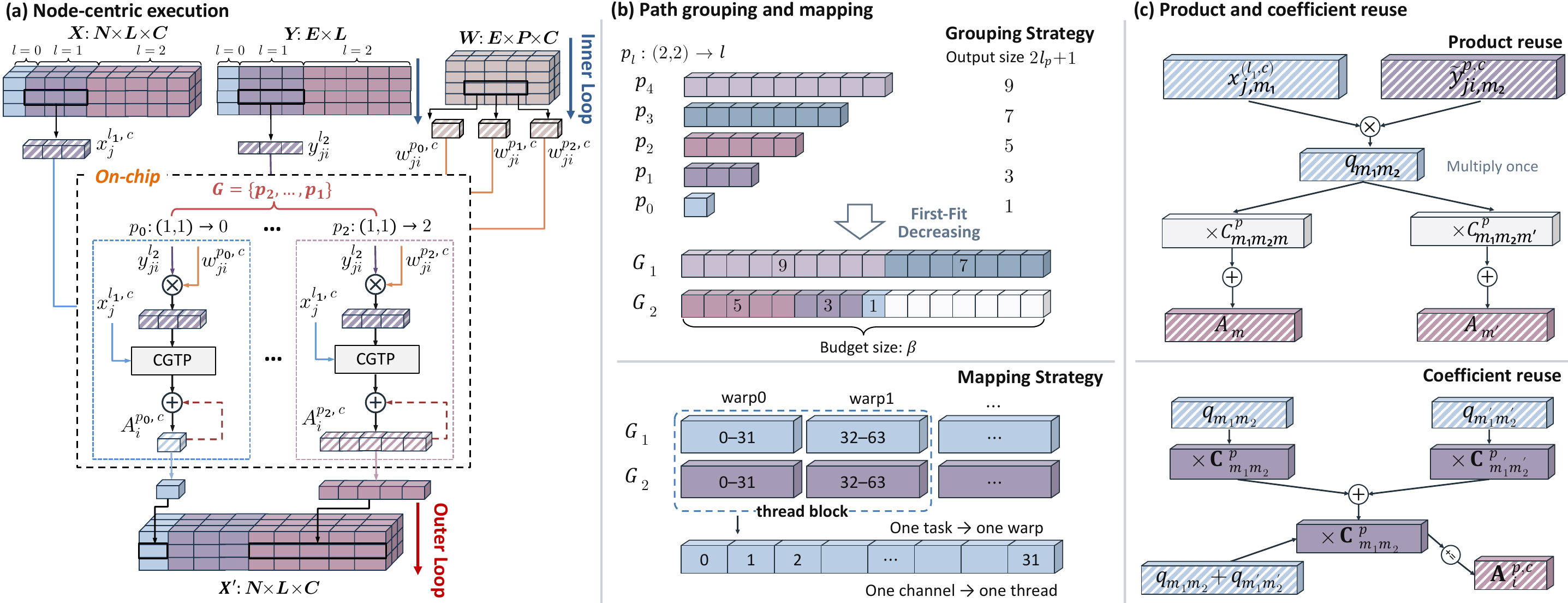}
    \caption{ 
    Overview of LunaTP.
    \textbf{(a)} Node-centric execution accumulates edge messages on chip before storing results.
    \textbf{(b)} Paths are grouped under an output component budget and mapped to warps by channel block.
    \textbf{(c)} Products are reused across outputs, and matching coefficients are factored.
    }
    \label{fig:lunatp-method}
\end{center}
\end{figure}
In this section, we introduce LunaTP’s node-centric execution (Section~\ref{sec: node-centric}). Building on this execution scheme, we describe path-group parallelism (Section~\ref{sec: partitioning}), followed by sparse expansion and multilevel reuse (Section~\ref{sec: reuse}). Finally, we present higher-order differentiation and GPU implementation details (Section~\ref{sec: implementation}).

\subsection{Node-Centric Execution}\label{sec: node-centric}
LunaTP organizes CGTP computation around destination nodes, accumulates edge messages on chip, and stores the results in global memory after the inner loop. Figure~\ref{fig:lunatp-method}(a) shows this execution scheme. The outer loop runs in parallel across destination nodes, paths, and channels, while the inner loop traverses the edges of each destination node. To support this traversal, LunaTP stores the adjacency list in CSR format by destination node, with the edges of node $i$ in the range $[\mathrm{rowptr}[i],\,\mathrm{rowptr}[i+1])$.

In each inner-loop iteration, a thread loads the source-node features $\mathbf{x}^{(l_1,c)}_j$, edge features $\mathbf{y}^{(l_2)}_{ji}$, and path weight $w^{p,c}_{ji}$ for edge $ji$. It computes CGTP between the source-node features and the weighted edge features $\widetilde{\mathbf{y}}^{p,c}_{ji}=w^{p,c}_{ji}\mathbf{y}^{(l_2)}_{ji}$. The thread computes only component products corresponding to nonzero CG terms, multiplies them by their CG coefficients, and accumulates the results directly in registers. For path $p=(l_1,l_2)\to l$, the accumulator $\mathbf{A}^{p}_{i,c}$ is defined as
\begin{equation}
\mathbf{A}^{p,c}_{i}
=
\sum_{j\in\mathcal{N}(i)}
\left[
\mathbf{x}^{(l_1,c)}_j
\otimes
\widetilde{\mathbf{y}}^{p,c}_{ji}
\right]^{(l)},
\label{eq:accumulator}
\end{equation}
where $\mathcal{N}(i)$ denotes the neighbors of node $i$, $w^{p,c}_{ji}$ is the path weight generated by the radial MLP, and the CG tensor product is defined in Eq.~\ref{eq:cgtp}.

After traversing all edges of the destination node, each thread stores its accumulated result once at the corresponding output location. Computation across channels and paths remains parallel, while global updates decrease from $O(E)$ to $O(N)$ for each fixed path and channel. However, the accumulators $\mathbf{A}^{p,c}_{i}$ need to remain in registers throughout the inner loop. Computing more paths that share inputs can improve reuse but also increases register demand. Work partitioning therefore needs to account for both input reuse and accumulator storage. The next section describes how path-group parallelism controls group size and organizes parallel computation.

\subsection{Path-Group Parallelism}\label{sec: partitioning}
Multiple paths for the same destination node can share source features. Prior work aggregates these paths to reduce repeated loads~\citep{lee2025flashtp}. However, the aggregation scope affects both reuse and register usage. Aggregating all paths that share inputs keeps their output accumulators live until the inner loop ends, which may cause register spilling and limit parallelism. Processing each path separately reduces register usage but repeatedly loads the same features, limiting input reuse. LunaTP therefore uses a grouping strategy to balance input reuse and register usage. For a path group $G$, let $l_p$ denote the output degree of path $p$. The number of output components retained per channel is
\begin{equation}
R_G=\sum_{p\in G}(2l_p+1),
\label{eq:register-budget}
\end{equation}
LunaTP uses $R_G\le\beta$ as the grouping constraint, where $\beta$ is the per-channel output component budget ($\beta=16$ in our experiments). We use First-Fit Decreasing (FFD) to group paths in descending order of output size, as shown in Figure~\ref{fig:lunatp-method}(b).

Since each path group can be computed independently, we use path groups as units of parallel work, which we call \emph{path-group parallelism}. Specifically, LunaTP partitions work across destination nodes, path groups, and channels. Each task is identified by $(i,G,B)$, where $i$ is the destination node, $G$ is the path group, and $B$ is a contiguous channel block. In the mapping strategy shown in Figure~\ref{fig:lunatp-method}(b), one warp processes one task covering $32$ consecutive channels, with each thread handling one channel and computing all paths in the group to reuse the loaded source features. Each task traverses all edges of its destination node and completes its accumulation on chip, avoiding an additional reduction to combine partial results from splitting the edge loop.

\paragraph{\textbf{Task ordering.}}
This partitioning provides more parallel tasks, but differences in node degree and per-path cost can still cause workload imbalance within a thread block. When some warps finish early, the block must wait for the remaining warps before releasing all its resources. To reduce this tail effect, LunaTP sorts path groups by their total number of nonzero CG terms in descending order, then sorts nodes within each group by degree in descending order, placing heavier tasks first and helping assign tasks with similar workloads to the same thread block.

\subsection{Sparse Expansion and Multi-Level Reuse}\label{sec: reuse}
LunaTP exploits sparsity and reuse in the CG structure to reduce memory accesses and computation. Path grouping enables input reuse across paths, while product and coefficient reuse further reduce computation within each path.

\paragraph{\textbf{Sparse expansion.}}
CG coefficients and their component indices are determined by the path and do not depend on the graph or input features. Like OpenEq~\citep{bharadwaj2025efficientsparsekernelgenerator}, LunaTP embeds normalized nonzero coefficients and their indices directly in the generated code, avoiding runtime table lookups and checks for zero terms.

Beyond CG sparsity, LunaTP uses path grouping to reuse loaded inputs across paths. In each inner-loop iteration, a thread loads each source-feature and edge-feature component needed by the group once and shares it across the paths that use it.

\paragraph{\textbf{Product reuse.}}
The same input product
$q_{m_1m_2}=x^{(l_1,c)}_{j,m_1}\widetilde{y}^{p,c}_{ji,m_2}$
may be used by multiple output components. LunaTP therefore organizes CG terms by input component pair, computes each product once, and reuses it for all terms with $C^p_{m_1m_2m}\neq 0$, as shown in the upper part of Figure~\ref{fig:lunatp-method}(c). Here, $m_1$ and $m_2$ follow the input component indices in Eq.~\ref{eq:cgtp}.

\paragraph{\textbf{Coefficient reuse.}}
Different input component pairs may have coefficient vectors that are identical up to an overall sign. Let $\mathbf{C}^{p}_{m_1m_2}\in\mathbb{R}^{2l+1}$ collect the CG coefficients over all output components. When
$\mathbf{C}^{p}_{m'_1m'_2}=\pm\mathbf{C}^{p}_{m_1m_2}$,
we combine the products before applying the shared coefficients,
\begin{equation}
\begin{aligned}
&q_{m_1m_2}\mathbf{C}^{p}_{m_1m_2}
+q_{m'_1m'_2}\mathbf{C}^{p}_{m'_1m'_2}
 =
(q_{m_1m_2}\pm q_{m'_1m'_2})
\mathbf{C}^{p}_{m_1m_2}.
\end{aligned}
\label{eq:signed-class}
\end{equation}
This reduces coefficient multiplications, as shown in the lower part of Figure~\ref{fig:lunatp-method}(c). In particular, we identify these matches only within each path at operator construction time.

\subsection{Implementation Details}\label{sec: implementation}

\paragraph{\textbf{Extension to higher-order derivatives.}}
LunaTP extends the node-centric execution scheme and reuse methods above to backward, double-backward, and triple-backward passes by adapting the reduction direction to the differentiation target. Computing the gradient of node features $\mathbf{x}_j$ requires accumulating results over all edges with source node $j$, so reduction shifts from destination nodes to source nodes. To support this traversal, LunaTP organizes a CSR by source node while retaining the partitioning across nodes, path groups, and channels and accumulating gradients on chip. Derivatives with respect to the incoming gradient instead reduce at destination nodes, as in the forward pass. Edge-feature and path-weight derivatives are computed per edge, with edge-feature derivatives also requiring a reduction across channels. Double-backward and triple-backward use the same approach, organizing computation around the reduction required by each output. The full pseudocode is provided in Appendix~\ref{app: derivation of tp conv}.

Higher-order backward passes also allow reuse across derivative branches. In double-backward, for example, computing node-feature and edge-feature derivatives requires the same incoming gradient for a given edge. LunaTP combines branches that share inputs to avoid repeated loads and applies sparse expansion, product reuse, and coefficient reuse within each branch.

\paragraph{\textbf{Low-level optimization and interface.}}
LunaTP uses \texttt{\_\_ldg} to load read-only features, weights, and adjacency indices, and assigns consecutive channels to threads within each warp to coalesce path-weight loads. We express accumulation as multiply-add operations so the compiler can generate fused multiply-add instructions. In higher-order backward passes, we assign independent derivative branches to different CUDA streams to overlap their execution. Meanwhile, we provide a programming interface for LunaTP to improve ease of use (see Appendix~\ref{app:programming-interface}).

\section{Experiments}\label{sec: experiments}

We evaluate LunaTP across four dimensions: computational efficiency, memory efficiency, scalability, and accuracy. Our baselines include e3nn \citep{geiger2022e3nneuclideanneuralnetworks}, FlashTP \citep{lee2025flashtp}, OpenEquivariance (OpenEq) \citep{bharadwaj2025efficientsparsekernelgenerator}, and cuEquivariance (cuEq) \citep{nvidia_cuequivariance}. 
We report kernel microbenchmark results (Section~\ref{sec:kernel bench}), end-to-end training and inference performance (Section~\ref{sec:end2end bench}), and scalability analysis (Section~\ref{sec:scalability analysis}). The experimental setup is detailed in Appendix~\ref{app: evaluation setup} and ablation studies are described in Appendix~\ref{sec: ablation}.

\subsection{Kernel Microbenchmark}\label{sec:kernel bench}
The microbenchmarks cover ten irrep configurations with $L_{\max}$ ranging from 2 to 6. Most are based on configurations used in existing equivariant MLIPs~\citep{Batzner_2022,batatia2022mace, Park_2024, kim_optimizing_2026, kavanagh2026fastaccuratefoundationmodels}. The hidden, edge, and output irreps and their representative MLIPs are summarized in Appendix Table~\ref{tab:kernel-workloads}.
\begin{figure}[h]
\begin{center}
    \includegraphics[width=1.0\linewidth]{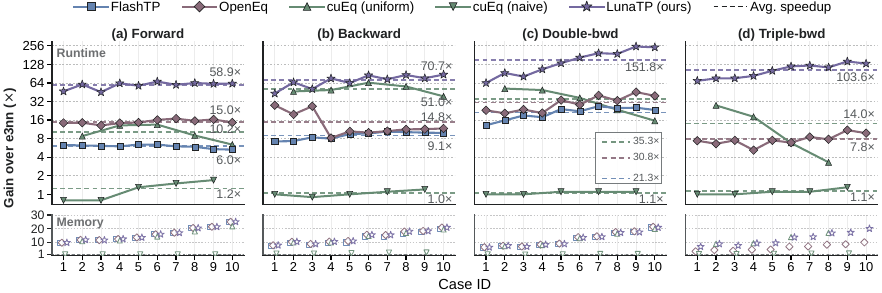}
    \caption{ 
        LunaTP performance in kernel microbenchmarks. Solid lines show runtime speedup over e3nn, while hollow markers show the reduction in peak memory relative to e3nn. Evaluations are performed in FP32 on NVIDIA H100 80GB GPUs. The raw data are provided in Appendix~\ref{app: raw data for kernel bench}.
    }
    \label{fig: kernel result}
\end{center}
\end{figure}

\paragraph{\textbf{Runtime and peak memory.}}
Figure~\ref{fig: kernel result} summarizes the FP32 microbenchmark results on H100 (FP64 results are provided in Appendix~\ref{sec:kernel result fp64}). LunaTP achieves average speedups of 58.9$\times$, 70.7$\times$, 151.8$\times$, and 103.6$\times$ over e3nn for forward, backward, double-backward, and triple-backward, respectively. Its corresponding speedups over OpenEq are 3.93$\times$, 5.93$\times$, 4.78$\times$, and 13.31$\times$. Although FlashTP and OpenEq already use kernel fusion to reduce memory traffic, LunaTP further improves performance through node-centric execution, path-group parallelism, and multilevel reuse. The gains are most pronounced for double- and triple-backward, which offer more opportunities to share data and computation across derivative branches. cuEq also performs well with uniform channel counts across degrees, but uses a naive implementation with limited speedups otherwise.

In addition to runtime improvements, LunaTP uses only 4.0\%--16.2\% of e3nn's peak memory. Its memory footprint is comparable to that of FlashTP and OpenEq from forward through double-backward, since all three avoid materializing per-edge message tensors. For triple-backward, LunaTP uses approximately 48\% of OpenEq's peak memory while also achieving lower runtime.

\paragraph{\textbf{Numerical stability and training convergence.}}
Since kernel fusion and parallel reductions can alter floating-point operation order, LunaTP and e3nn may exhibit small numerical discrepancies. Using e3nn's FP64 results as the reference, we measure the RMSEs of their FP32 outputs and corresponding derivatives. From the forward pass through third-order derivatives, both implementations maintain FP32 RMSEs on the order of $10^{-7}$--$10^{-6}$, while LunaTP's FP64 RMSEs remain between $10^{-16}$ and $10^{-14}$ (see Appendix~\ref{app: numerical stability}). 

To assess training stability, we train NequIP-XL \citep{Batzner_2022, kavanagh2026fastaccuratefoundationmodels} on MPTrj-PFT \citep{deng_2023_chgnet, koker2026pft} with the same settings. LunaTP’s validation MAEs for E, F, S, and H differ from those of e3nn by less than 0.3\% across all targets, while showing similar convergence trends. (see Appendix~\ref{app: training convergence}).

\paragraph{\textbf{Hardware portability.}}
We evaluate LunaTP on NVIDIA A100 and RTX 3090 GPUs to assess its portability across data-center and consumer hardware. These GPUs differ in compute performance and memory bandwidth. LunaTP also has a performance advantage over the baselines on both GPUs. Full results are reported in Appendix~\ref{sec:kernel result portability}.

\subsection{End-to-end Performance}\label{sec:end2end bench}
We evaluate LunaTP's end-to-end performance in training and inference. For training, we measure throughput on eight state-of-the-art (SOTA) equivariant MLIPs with EFS and EFSH objectives. For inference, we measure MD speed and the maximum system size each backend can run. Experimental details are in Appendix~\ref{app: end2end settings}.
\begin{figure}[t]
\begin{center}
    \includegraphics[width=1.0\linewidth]{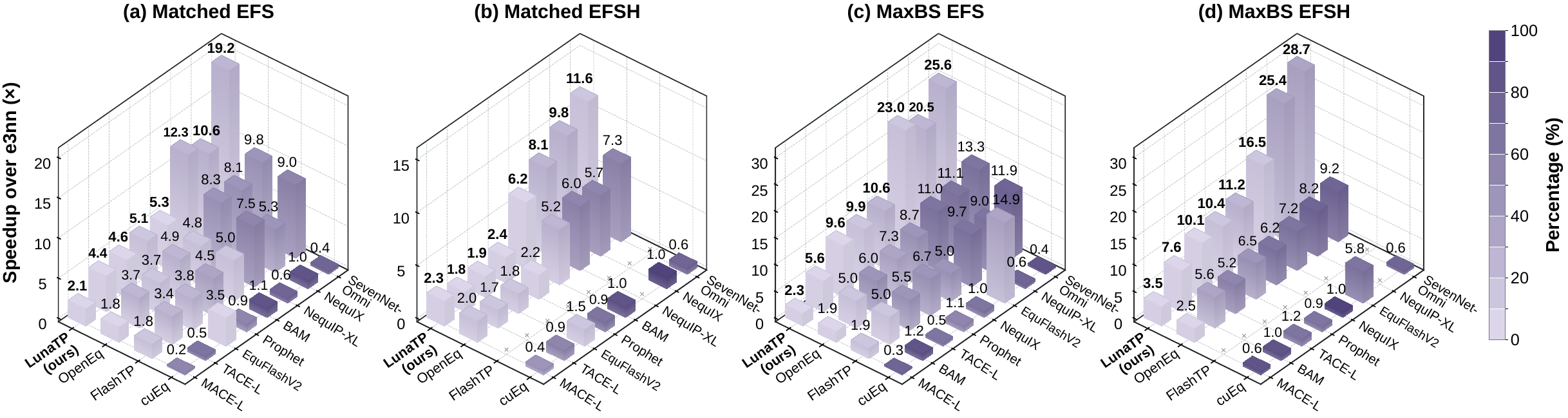}
    \caption{ 
    End-to-end training performance.
    We benchmark EFS (energy, forces, and stress) and EFSH (EFS plus a Hessian objective). Matched uses equal batch sizes across backends, and MaxBS selects each backend’s maximum batch size. Bars show throughput speedup over e3nn, while bar color indicates the CGTP convolution time fraction.
    The raw data are provided in Appendix~\ref{app: raw data for training}.
    }
    \label{fig: training result}
\end{center}
\end{figure}

\paragraph{\textbf{Training.}}
As shown in Figure~\ref{fig: training result}(a) and (b), LunaTP achieves speedups of 2.10--19.24$\times$ over e3nn for EFS and 1.81--11.6$\times$ for EFSH at matched batch sizes (Matched). The speedup depends on the fraction of training time spent in CGTP convolutions. For example, MACE-L achieves a 2.10$\times$ speedup under Matched EFS. With e3nn, CGTP convolutions account for 53.5\% of runtime, limiting the overall speedup achievable by optimizing this operator. In contrast, SevenNet-Omni spends a larger fraction of runtime in CGTP convolutions and achieves a 19.2$\times$ speedup. Under Matched EFSH, LunaTP achieves speedups of 1.06--1.72$\times$ over OpenEq, with further gains also limited by work outside CGTP convolutions.

To evaluate larger workloads, we increase the batch size to the maximum each backend can run on a single GPU (MaxBS). As shown in Figure~\ref{fig: training result}(c) and (d), LunaTP achieves speedups of 2.30--25.6$\times$ over e3nn for EFS and 3.50--28.7$\times$ for EFSH. The corresponding speedups over OpenEq are 1.14--2.10$\times$ and 1.34--3.11$\times$. These gains over OpenEq generally exceed those under Matched, showing that LunaTP provides greater end-to-end benefits on larger workloads.

\paragraph{\textbf{Inference.}}
\begin{wrapfigure}[17]{r}{0.45\textwidth}
    \centering
    \includegraphics[width=0.85\linewidth]{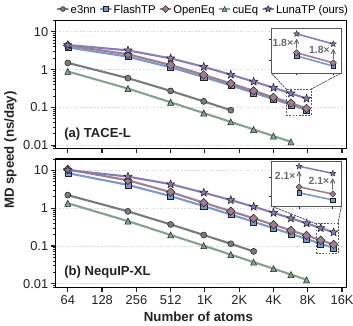}
    \caption{
    Inference scaling. MD speed on Si with a lattice of 5.43~\AA\ and a 1-fs time step.
    }
    \label{fig: inference result}
\end{wrapfigure}
We evaluate inference scalability on TACE-L and NequIP-XL, as shown in Figure~\ref{fig: inference result}. LunaTP achieves an 8.70$\times$ speedup on TACE-L and a 15.27$\times$ speedup on NequIP-XL at e3nn's maximum system size. 
Since kernel fusion avoids storing per-edge message tensors, LunaTP increases the maximum atom count by 4.63$\times$ and 5.04$\times$ for TACE-L and NequIP-XL, respectively.
OpenEq and FlashTP also use kernel fusion, so all three backends reach the same maximum system size. At these workloads, LunaTP remains 1.8$\times$ faster on TACE-L and 2.1$\times$ faster on NequIP-XL than OpenEq. This shows that LunaTP's optimizations further improve inference performance even with kernel fusion already in place.

\subsection{Scalability Analysis}\label{sec:scalability analysis}
\begin{figure}[h]
\begin{center}
    \includegraphics[width=1.0\linewidth]{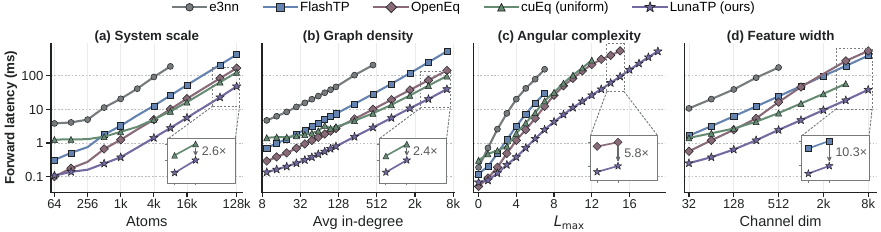}
    \captionsetup{justification=centering,singlelinecheck=false}
    \caption{ 
        Forward scalability of CGTP convolution.
    }
    \label{fig: kernel scaling}
\end{center}
\end{figure}
We evaluate LunaTP's forward scalability across four dimensions: atom count, average in-degree, maximum degree, and channel size. As Figure~\ref{fig: kernel scaling} shows, latency generally increases with workload size, but LunaTP achieves the lowest latency at nearly all measured points. At the largest comparable point in each sweep, it is 2.4--10.3$\times$ faster than the fastest available baseline, with larger advantages in the maximum-degree and channel-size sweeps. As $L_{\max}$ increases, the number of coupling paths and CG terms grows. LunaTP controls accumulator storage through path grouping and reduces repeated computation through multilevel reuse. As channels become wider, LunaTP partitions the additional computation into more independent channel-block tasks that run across warps without increasing the per-thread accumulator count under the basic mapping. These results demonstrate LunaTP's scalability to large graphs, higher-degree features, and wider channels. The experimental setup is provided in Appendix~\ref{sec:app scalabilty settings}.

\section{Related Work}\label{sec: related work}
\paragraph{\textbf{Algorithm-level optimization.}}
SO(2)-based methods align a local coordinate system with each edge direction to reduce SO(3) convolutions to simpler SO(2) operations~\citep{3618408.3619548}. Other approaches reformulate feature representations and coupling operations to exploit efficient computational primitives. Fused Tensor Product organizes computations across irreps into matrix multiplications~\citep{unke2024e3xmathrme3equivariantdeeplearning}, while Gaunt Tensor Product uses a Fourier basis and FFTs to accelerate products of spherical functions~\citep{luo2024enabling}. These approaches offer efficient alternatives for high-degree equivariant models, although their parameterizations can involve trade-offs between expressivity and efficiency~\citep{xie2025the}. LunaTP instead improves execution efficiency while preserving the original CGTP formulation and expressivity.

\paragraph{\textbf{Operator-level optimization.}}
A complementary direction improves tensor product implementations on GPUs. e3nn provides general-purpose equivariant operators~\citep{geiger2022e3nneuclideanneuralnetworks}, while cuEquivariance offers specialized GPU kernels~\citep{nvidia_cuequivariance}. FlashTP exploits CG sparsity, aggregates paths sharing source features, and fuses convolution operations to reduce repeated loads and intermediate storage~\citep{lee2025flashtp}. OpenEq generates sparse kernels just in time for CG contractions and supports fused graph convolutions and higher-order derivatives. Its deterministic implementation accumulates over sorted edges to reduce global writes, then merges node partial sums across warps with a fixup kernel
~\citep{bharadwaj2025efficientsparsekernelgenerator}. Building on these advances, LunaTP focuses on parallel execution and reuse within node-centric computation. It groups paths under an accumulator budget to expose intra-node parallelism and reuses common products and coefficients to reduce computation. Our experiments demonstrate performance gains over these baselines.

\section{Conclusion}\label{sec: conclusion}
In this work, we rethink the parallelism of CGTP convolutions and present LunaTP, a GPU kernel for CGTP convolutions in equivariant MLIPs. By organizing node reduction, data reuse, and work partitioning, LunaTP improves computational efficiency without sacrificing expressivity, and extends to third-order derivatives, making efficient Hessian-based training practical. Experiments show that LunaTP accelerates kernels by up to $250.0\times$ over e3nn and $17.1\times$ over OpenEq, while delivering up to $28.7\times$ the end-to-end training throughput of e3nn. Future work will extend these optimizations to other operators in equivariant MLIPs.

\subsection*{AI use statement}
In this work, we did not use generative AI tools for any tasks requiring disclosure. We used generative AI tools only for language editing to improve the clarity and readability of the manuscript and for literature search, which are tasks with recommended disclosure. We reviewed all AI-assisted edits and independently verified the identified literature and references. We take responsibility for the final content of this work, including all AI-assisted language edits and literature-related content.

\subsection*{Reproducibility statement}
We describe our methods in Section~\ref{sec: methods}. Appendix~\ref{app: evaluation setup} provides the experimental setup, including the hardware and software environments, benchmarking protocols, and model configurations. The raw benchmark data underlying the results reported in the main text are provided in Appendix~\ref{app: detailed benchmark data}. We will make the complete source code publicly available. Our code is available at \url{https://github.com/HPC-AI-Team/LunaEquivariance}

\subsection*{Acknowledgments}
This work is supported by the following funding: National Science Foundation of China (92270206, 62372435, T2125013, 62502501), Beijing Natural Science Foundation (4254087), National Key Research and Development Program of China (2025YFB4507000), and the Innovation Funding of ICT, CAS. Part of the training is performed on the robotic AI-Scientist platform of Chinese Academy of Sciences.

\bibliography{conference}
\bibliographystyle{plainnat}

\newpage
\appendix
\section*{Appendix}

\begin{itemize}[itemsep=10pt, parsep=5pt, topsep=10pt]

    \item[\ref{app: evaluation setup}] Evaluation setup
    \begin{itemize}[itemsep=8pt, parsep=4pt, topsep=8pt]
        %\item[\ref{sec:app profiling settings}] \textcolor{red}{Profiling workload settings}
        \item[\ref{sec:app evalution settings}] Kernel microbenchmark settings
        \item[\ref{sec:app scalabilty settings}] Scalability analysis settings
        \item[\ref{app: end2end settings}] Training setup details
    \end{itemize}
    \item[\ref{app: detailed benchmark data}] Detailed Benchmark Data
    \begin{itemize}[itemsep=8pt, parsep=4pt, topsep=8pt]
        \item[\ref{app: raw data for kernel bench}] Raw data for kernel microbenchmarks
        \item[\ref{app: raw data for training}] Raw data for training benchmarks
    \end{itemize}
    \item[\ref{sec: ablation}] Ablation Studies
    \begin{itemize}[itemsep=8pt, parsep=4pt, topsep=8pt]
        \item[\ref{sec: path grouping}] Path grouping
        \item[\ref{sec: task ordering}] Task ordering
        \item[\ref{sec: compile-time reuse}] Compile-time reuse
    \end{itemize}
    \item[\ref{app: addition kernel bench}] Additional kernel microbenchmark results
    \begin{itemize}[itemsep=8pt, parsep=4pt, topsep=8pt]
        \item[\ref{sec:kernel result fp64}] Kernel microbenchmark (FP64)
        \item[\ref{sec:kernel result portability}] Kernel microbenchmark (A100 \& RTX 3090)
    \end{itemize}

    \item[\ref{app:numerical stability and training convergence}] Details on numerical stability and training convergence
    \begin{itemize}[itemsep=8pt, parsep=4pt, topsep=8pt]
        \item[\ref{app: numerical stability}] Operator-level numerical stability
        \item[\ref{app: training convergence}]End-to-end training convergence
    \end{itemize}
    \item[\ref{app: derivation of tp conv}] Differentiation of CGTP convolution
    \item[\ref{app:programming-interface}] Programming interface
\end{itemize}

\clearpage

\section{Evaluation setup}\label{app: evaluation setup}
\paragraph{\textbf{Software and hardware.}}\label{app: software and hardware}
All evaluations use FP32 with \texttt{PyTorch 2.11.0+cu128}, e3nn v0.6.0 \citep{geiger2022e3nneuclideanneuralnetworks}, FlashTP v0.1.1 \citep{lee2025flashtp}, OpenEquivariance v0.7.0 \citep{bharadwaj2025efficientsparsekernelgenerator}, and cuEquivariance v0.11.1 \citep{nvidia_cuequivariance}. Notably, cuEquivariance uses \texttt{uniform\_1d} when all segments have the same shape and \texttt{naive} otherwise, which we denote as cuEq(uniform) and cuEq(naive), respectively.

Unless otherwise specified, all experiments are conducted on NVIDIA H100-80GB SXM GPUs with Intel Xeon Platinum 8468 CPUs. NVLink provides up to 900 GB/s of bidirectional bandwidth per GPU.

\paragraph{\textbf{Workload construction protocol.}}
%为了贴近真实的原子数据分布，我们随机打乱 OMat24 dataset \citep{barros-luque_open_2026} ，再随机选取完整结构组成具有指定原子数的批次。每个结构均在周期性边界条件下使用 $r_{\mathrm{cut}}=6.0~\text{\AA}$ 构建邻居图，从而保留数据集的实际原子邻居分布。
To construct workloads that reflect realistic atomic data distributions, we randomly shuffle the OMat24 dataset \citep{barros-luque_open_2026} and sample structures to form batches with specified atom counts. For each structure, we construct a neighbor graph under periodic boundary conditions with $r_{\mathrm{cut}}=6.0~\text{\AA}$, thereby preserving the neighbor distribution of the original dataset.

\paragraph{\textbf{Measurement protocol.}}\label{app: measurement protocol}
%对于所有算子级实验，每个配置在正式计时前预热 10 次，随后使用 CUDA events 重复测量 20 次，并报告平均运行时间。首次 JIT 编译、图构建和主机到设备的数据传输不计入。
For all microbenchmarks, we warm up each configuration for 10 iterations and then measure GPU kernel execution time over the next 20 iterations using the PyTorch profiler. We report the mean runtime, excluding initial JIT compilation, graph construction, and host-to-device data transfers.

%\subsection{Profiling Workload Settings}\label{sec:app profiling settings}
%按照上述工作负载构建流程，第~\ref{sec:performance-analysis}节的性能分析采用包含 768 个节点和 37,304 条有向边的工作负载，每个原子平均具有 48.6 个邻居。该工作负载采用基于 NequIP-OAM-L \citep{kavanagh2026fastaccuratefoundationmodels} 的 irrep 配置。隐藏、边和输出 irreps 分别为 \texttt{128x0e + 64x1e + 32x2e + 32x3e}、\texttt{1x0e + 1x1e + 1x2e + 1x3e} 和 \texttt{256x0e + 480x1e + 544x2e + 480x3e}。
%\textcolor{red}{Following the workload construction protocol described above, the performance analysis in Section~\ref{sec: performance analysis} uses a workload with 768 nodes and 37304 edges, corresponding to 48.6 neighbors per atom on average. The workload uses an irrep configuration based on NequIP-OAM-L \citep{kavanagh2026fastaccuratefoundationmodels}. The hidden, edge, and output irreps are \texttt{128x0e + 64x1e + 32x2e + 32x3e}, \texttt{1x0e + 1x1e + 1x2e + 1x3e}, and \texttt{256x0e + 480x1e + 544x2e + 480x3e}, respectively.}

\subsection{Kernel Microbenchmark Settings}\label{sec:app evalution settings}
%第~\ref{sec}节中的所有 kernel 微基准测试均采用相同的工作负载。除非特殊说明，否则测试的工作负载包含 768 个节点和 37,304 条有向边，平均每个原子具有 48.6 个邻居。为覆盖不同的角动量阶数和通道规模，我们选取了多个 SOTA 等变 MLIP 使用的 irrep 配置，具体配置汇总于表~\ref{tab:kernel-workloads}。
% 【Case id | L_max | Hidden irreps | edge irreps | output irreps | 代表性MLIP】
All kernel microbenchmarks in Section~\ref{sec:kernel bench} use the same workload unless otherwise specified. The workload contains 768 nodes and 37304 edges, with an average of 48.6 neighbors per atom. To cover different angular-momentum orders and channel sizes, we use irrep configurations from several SOTA equivariant MLIPs, as summarized in Table~\ref{tab:kernel-workloads}.
\begin{table*}[h]
\centering
\captionsetup{justification=centering,singlelinecheck=false}
\caption{Irrep configurations used in the kernel microbenchmarks.}
\label{tab:kernel-workloads}
\small
\setlength{\tabcolsep}{4pt}
\resizebox{\textwidth}{!}{
\begin{tabular}{cclllc}
\toprule
Case ID & $L_{\max}$ & Hidden irreps & Edge irreps & Output irreps & Representative MLIP \\
\midrule

1 & 2 &
\makecell[l]{128x0e+64x1e+32x2e} &
\makecell[l]{1x0e+1x1e\\+1x2e} &
\makecell[l]{224x0e+384x1e\\+352x2e} &
\makecell[c]{SevenNet-0 ,\\GNoME} \\
\midrule

2 & 2 &
\makecell[l]{128x0e+128x1o+128x2e} &
\makecell[l]{1x0e+1x1o\\+1x2e+1x3o} &
\makecell[l]{384x0e+640x1o\\+640x2e+512x3o} &
\makecell[c]{MACE-MP-L} \\
\midrule

3 & 3 &
\makecell[l]{128x0e+64x1o+32x2e\\+32x3o} &
\makecell[l]{1x0e+1x1o\\+1x2e+1x3o} &
\makecell[l]{256x0e+352x1o\\+384x2e+320x3o} &
\makecell[c]{Nequix-MP/OAM ,\\SevenNet-MF/Omni} \\
\midrule

4 & 3 &
\makecell[l]{256x0e+256x1o+256x2e\\+256x3o} &
\makecell[l]{1x0e+1x1o\\+1x2e+1x3o} &
\makecell[l]{1024x0e+1536x1o\\+1792x2e+1536x3o} &
EquFlash-OMat/OAM \\
\midrule

5 & 4 &
\makecell[l]{320x0e+96x1o+64x2e\\+32x3o+32x4e} &
\makecell[l]{1x0e+1x1o+1x2e\\+1x3o+1x4e} &
\makecell[l]{544x0e+736x1o+832x2e\\+800x3o+704x4e} &
NequIP-XL series \\
\midrule

6 & 4 &
\makecell[l]{128x0e+128x1o+128x2e\\+128x3o+128x4e} &
\makecell[l]{1x0e+1x1o+1x2e\\+1x3o+1x4e} &
\makecell[l]{640x0e+1024x1o+1280x2e\\+1280x3o+1152x4e} &
EquFlashV2-OMat/OAM \\
\midrule

7 & 5 &
\makecell[l]{128x0e+64x1o+32x2e\\+32x3o+32x4e+32x5o} &
\makecell[l]{1x0e+1x1o+1x2e\\+1x3o+1x4e+1x5o} &
\makecell[l]{320x0e+480x1o+576x2e\\+608x3o+608x4e+512x5o} &
-- \\
\midrule

8$^*$ & 5 &
\makecell[l]{96x0e+96x1o+96x2e\\+96x3o+96x4e+96x5o} &
\makecell[l]{1x0e+1x1o+1x2e\\+1x3o+1x4e+1x5o} &
\makecell[l]{576x0e+960x1o+1248x2e\\+1344x3o+1344x4e+1152x5o} &
-- \\
\midrule

9 & 6 &
\makecell[l]{128x0e+64x1o+32x2e+32x3o\\+32x4e+32x5o+32x6e} &
\makecell[l]{1x0e+1x1o+1x2e+1x3o\\+1x4e+1x5o+1x6e} &
\makecell[l]{352x0e+544x1o+672x2e+736x3o\\+768x4e+736x5o+640x6e} &
-- \\
\midrule

10$^*$ & 6 &
\makecell[l]{64x0e+64x1o+64x2e+64x3o\\+64x4e+64x5o+64x6e} &
\makecell[l]{1x0e+1x1o+1x2e+1x3o\\+1x4e+1x5o+1x6e} &
\makecell[l]{448x0e+768x1o+1024x2e+1152x3o\\+1216x4e+1152x5o+1024x6e} &
-- \\

\bottomrule
\end{tabular}
}

\vspace{1.0pt}
    \parbox{\textwidth}
    {\footnotesize 
        ``*'' e3nn encounters OOM under the default workload; these cases therefore use 448 atoms and 20386 edges. \\
        ``--'' Since no existing MLIP uses the corresponding settings, we constructed these configurations for evaluation.
    }
\end{table*}

\subsection{Scalability Analysis Settings}\label{sec:app scalabilty settings}
We evaluate LunaTP's FP32 forward scalability along four dimensions: atom count, graph density, maximum degree, and channel size. In each test, we vary one dimension while holding the others fixed, and continue until OOM or compilation timeout.
\paragraph{\textbf{Scaling with atom count.}}
We combine distinct OMat structures and double the atom count starting from 16. We use a 6.0~\AA\ neighbor cutoff, $L_{\max}=4$, and 64 channels per degree.
\paragraph{\textbf{Scaling with graph density.}}
Using a fixed 768-atom batch, we increase the neighbor cutoff from 3.5~\AA\ to 30.5~\AA\ while keeping $L_{\max}=4$ and 64 channels per degree.
\paragraph{\textbf{Scaling with degree.}}
For this test and the channel-size test, we use the same fixed 768-atom graph. We hold the channel count at 64 per degree and increase $L_{\max}$ from 0 in steps of one.
\paragraph{\textbf{Scaling with channel size.}}
We fix the 768-atom graph and $L_{\max}=4$, and increase the number of channels per degree from 32 by successive doubling.

\subsection{Training Setup Details}\label{app: end2end settings}
%我们在NVIDIA H100 GPU中，FP32精度下训练，其中EFS测试采用OMat-1M\citep{OMat},EFSH测试采用MPTrj-PFT\citep{deng, pft}数据集。所有模型基于各自官方实现，并且采用torch进行训练。特别的，在EFSH训练中，我们采用\citet{pft}的训练流程，对hessian矩阵进行Column采样。
We benchmark training in FP32 on NVIDIA H100 GPUs, using OMat-1M~\citep{barros-luque_open_2026} for EFS and MPTrj-PFT~\citep{deng_2023_chgnet, koker2026pft} for EFSH. All models are trained using their official PyTorch implementations~\citep{10.1063/5.0297006,kim_optimizing_2026,koker2026pft,xu2026spectralspatialtensoratomiccluster,kavanagh2026fastaccuratefoundationmodels,willow2026bayesiane3equivariantinteratomicpotential}. For EFSH, we follow the training protocol of \citet{koker2026pft} and sample Hessian columns during training. Tables~\ref{tab:training-batch-settings} and~\ref{tab:training-model-settings} summarize per-GPU batch sizes and the model configurations, respectively.

\begin{table}[h]
\centering
\caption{Training batch sizes (atoms per GPU). Matched uses the same batch size across all backends, while MaxBS uses the maximum batch size supported by each backend.}
\label{tab:training-batch-settings}
\scriptsize
\setlength{\tabcolsep}{3.0pt}
\renewcommand{\arraystretch}{1.16}

\begin{minipage}{1.0\textwidth}
\centering
\begin{tabular*}{\linewidth}{@{\extracolsep{\fill}}l|cccccccc@{}}
\toprule
\multicolumn{1}{c|}{\textbf{Atoms}} & \textbf{SevenNet-Omni} & \textbf{MACE-L} &
\textbf{NequIP-XL} & \textbf{EquFlashV2} & \textbf{TACE-L} & \textbf{NequIX} & \textbf{Prophet} & \textbf{BAM}\\
\midrule
\multicolumn{9}{c}{\textit{Matched}} \\
EFS  & 2340 & 2450 & 1630 & 290 & 1030 & 8900 & 430 & 1280 \\
EFSH &  870 &  800 &  600 &  90 &  300 & 3200 & 120 & 580 \\
\midrule
\multicolumn{9}{c}{\textit{MaxBS: EFS}} \\
e3nn    &  2340 & 2450 & 1630 &  290 & 1030 & 8900 & 430 & 1280 \\
FlashTP & 14600 & 3850 & 9550 & 2550 & 3940 & 30400 & 2220 & 3440\\
OpenEq  & 14600 & 3850 & 9550 & 2550 & 3940 & 37600 & 2220 & 3440\\
cuEq    & 14600 & 3850 & 3300 & 2550 & 2340 & 14500 & 510 & 1600 \\
LunaTP  & 14600 & 3850 & 9550 & 2550 & 3940 & 37600 & 2220 & 3440\\
\midrule
\multicolumn{9}{c}{\textit{MaxBS: EFSH}} \\
e3nn    &  870 &  800 &  600 &  90 & 300 & 3200 & 120 & 580 \\
FlashTP & -- & -- & -- & -- & -- & -- & -- & -- \\
OpenEq  & 3830 & 2600 & 2750 & 660 & 1380 & 9870 & 600 & 1700 \\
cuEq    & 3830 & 3140 & -- & 660 & 1340 & 3060 & 120 & 580 \\
LunaTP  & 5130 & 6200 & 3730 & 860 & 1650 & 13200 & 680 & 1700 \\
\bottomrule
\end{tabular*}
\end{minipage}

\end{table}

\begin{table}[h]
\centering
\captionsetup{justification=centering,singlelinecheck=false}
\caption{MLIP configurations for training.}
\label{tab:training-model-settings}
\scriptsize
\setlength{\tabcolsep}{2.0pt}
\renewcommand{\arraystretch}{1.16}

\begin{subtable}[t]{0.320\textwidth}
\centering
\caption{MACE-L (10.10M)}
\begin{tabular*}{\linewidth}{@{\extracolsep{\fill}}>{\raggedright\arraybackslash}p{0.49\linewidth}|>{\raggedright\arraybackslash}p{0.425\linewidth}@{}}
\toprule
\multicolumn{1}{c|}{\textbf{Hyper-parameter}} & \multicolumn{1}{c}{\textbf{Value}} \\
\midrule
Layers & 2 \\
Cutoff radius & 5.0$~\text{\AA}$ \\
\midrule
\multicolumn{1}{c|}{\textbf{Stage}} & \multicolumn{1}{c}{\textbf{Irreps}} \\
\midrule
Edge & \makecell[l]{1x0e+1x1o\\+1x2e+1x3o} \\
\arrayrulecolor{white}\midrule
Input / Output(L2) & 128x0e / 128x0e \\
\arrayrulecolor{white}\midrule
Hidden (L1) & \makecell[l]{128x0e\\+128x1o\\+128x2e} \\
\arrayrulecolor{black}
\bottomrule
\end{tabular*}
\end{subtable}\hfill
\begin{subtable}[t]{0.320\textwidth}
\centering
\caption{NequIP-XL (32.14M)}
\begin{tabular*}{\linewidth}{@{\extracolsep{\fill}}>{\raggedright\arraybackslash}p{0.44\linewidth}|>{\raggedright\arraybackslash}p{0.475\linewidth}@{}}
\toprule
\multicolumn{1}{c|}{\textbf{Hyper-parameter}} & \multicolumn{1}{c}{\textbf{Value}} \\
\midrule
Layers & 6 \\
Cutoff radius & 6.0$~\text{\AA}$ \\
\midrule
\multicolumn{1}{c|}{\textbf{Stage}} & \multicolumn{1}{c}{\textbf{Irreps}} \\
\midrule
Edge & \makecell[l]{1x0e+1x1o+1x2e\\+1x3o+1x4e} \\
\arrayrulecolor{white}\midrule
Input / Output(L6) & 32x0e / 320x0e \\
\arrayrulecolor{white}\midrule
Hidden (L1--5) & \makecell[l]{320x0e+96x1o\\+64x2e+32x3o\\+32x4e} \\
\arrayrulecolor{black}
\bottomrule
\end{tabular*}
\end{subtable}\hfill
\begin{subtable}[t]{0.320\textwidth}
\centering
\caption{EquFlashV2 (44.89M)}
\begin{tabular*}{\linewidth}{@{\extracolsep{\fill}}>{\raggedright\arraybackslash}p{0.44\linewidth}|>{\raggedright\arraybackslash}p{0.475\linewidth}@{}}
\toprule
\multicolumn{1}{c|}{\textbf{Hyper-parameter}} & \multicolumn{1}{c}{\textbf{Value}} \\
\midrule
Layers & 10 \\
Cutoff radius & 6.0$~\text{\AA}$ \\
\midrule
\multicolumn{1}{c|}{\textbf{Stage}} & \multicolumn{1}{c}{\textbf{Irreps}} \\
\midrule
Edge & \makecell[l]{1x0e+1x1o+1x2e\\+1x3o+1x4e} \\
\arrayrulecolor{white}\midrule
Input / Output(L10) & 128x0e / 128x0e \\
\arrayrulecolor{white}\midrule
Hidden (L1--9) & \makecell[l]{128x0e+128x1o\\+128x2e+128x3o\\+128x4e} \\
\arrayrulecolor{black}
\bottomrule
\end{tabular*}
\end{subtable}

\par\vspace{1.3em}

\begin{subtable}[t]{0.320\textwidth}
\centering
\caption{TACE-L (82.90M)}
\begin{tabular*}{\linewidth}{@{\extracolsep{\fill}}>{\raggedright\arraybackslash}p{0.44\linewidth}|>{\raggedright\arraybackslash}p{0.475\linewidth}@{}}
\toprule
\multicolumn{1}{c|}{\textbf{Hyper-parameter}} & \multicolumn{1}{c}{\textbf{Value}} \\
\midrule
Layers & 5 \\
Cutoff radius & 6.0$~\text{\AA}$ \\
\midrule
\multicolumn{1}{c|}{\textbf{Stage}} & \multicolumn{1}{c}{\textbf{Irreps}} \\
\midrule
Edge & \makecell[l]{1x0e+1x1o+1x2e\\+1x3o+1x4e+1x5o} \\
\arrayrulecolor{white}\midrule
\makecell[l]{Input / Output (L5)} & \makecell[l]{64x0e / 64x0e}\\
\arrayrulecolor{white}\midrule
Hidden (L1--4) & \makecell[l]{64x0e+64x1o\\+64x2e+64x3o\\+64x4e+64x5o} \\
\arrayrulecolor{black}
\bottomrule
\end{tabular*}
\end{subtable}\hfill
\begin{subtable}[t]{0.320\textwidth}
\centering
\caption{NequIX (0.71M)}
\begin{tabular*}{\linewidth}{@{\extracolsep{\fill}}>{\raggedright\arraybackslash}p{0.44\linewidth}|>{\raggedright\arraybackslash}p{0.475\linewidth}@{}}
\toprule
\multicolumn{1}{c|}{\textbf{Hyper-parameter}} & \multicolumn{1}{c}{\textbf{Value}} \\
\midrule
Layers & 4 \\
Cutoff radius & 6.0$~\text{\AA}$ \\
\midrule
\multicolumn{1}{c|}{\textbf{Stage}} & \multicolumn{1}{c}{\textbf{Irreps}} \\
\midrule
Edge & \makecell[l]{1x0e+1x1o\\+1x2e+1x3o} \\
\arrayrulecolor{white}\midrule
\makecell[l]{Input / Output (L4)} & \makecell[l]{89x0e / 128x0e} \\
\arrayrulecolor{white}\midrule
Hidden (L1--3) & \makecell[l]{128x0e+64x1o\\+32x2e\\+32x3o} \\
\arrayrulecolor{black}
\bottomrule
\end{tabular*}
\end{subtable}\hfill
\begin{subtable}[t]{0.320\textwidth}
\centering
\caption{Prophet (62.30M)}
\begin{tabular*}{\linewidth}{@{\extracolsep{\fill}}>{\raggedright\arraybackslash}p{0.44\linewidth}|>{\raggedright\arraybackslash}p{0.475\linewidth}@{}}
\toprule
\multicolumn{1}{c|}{\textbf{Hyper-parameter}} & \multicolumn{1}{c}{\textbf{Value}} \\
\midrule
Layers & 10 \\
Cutoff radii & 3.0, 7.0$~\text{\AA}$ \\
\midrule
\multicolumn{1}{c|}{\textbf{Stage}} & \multicolumn{1}{c}{\textbf{Irreps}} \\
\midrule
Edge & \makecell[l]{1x0e+1x1o+1x2e\\+1x3o+1x4e} \\
\arrayrulecolor{white}\midrule
\makecell[l]{Input / Output(L10)} & \makecell[l]{256x0e / 256x0e} \\
\arrayrulecolor{white}\midrule
Hidden (L1--9) & \makecell[l]{256x0e+96x1o\\+64x2e+32x3o\\+32x4e} \\
\arrayrulecolor{black}
\bottomrule
\end{tabular*}
\end{subtable}

\par\vspace{1.3em}

\begin{subtable}[t]{0.460\textwidth}
\centering
\caption{SevenNet-Omni (25.76M)}
\begin{tabular*}{\linewidth}{@{\extracolsep{\fill}}>{\raggedright\arraybackslash}p{0.43\linewidth}|>{\raggedright\arraybackslash}p{0.52\linewidth}@{}}
\toprule
\multicolumn{1}{c|}{\textbf{Hyper-parameter}} & \multicolumn{1}{c}{\textbf{Value}} \\
\midrule
Layers & 5 \\
Cutoff radius & 6.0$~\text{\AA}$ \\
\midrule
\multicolumn{1}{c|}{\textbf{Stage}} & \multicolumn{1}{c}{\textbf{Irreps}} \\
\midrule
Edge & 1x0e+1x1e+1x2e+1x3e \\
\arrayrulecolor{white}\midrule
Input / Output (L5) & 128x0e / 128x0e \\
\arrayrulecolor{white}\midrule
Hidden (L1, L4) & \makecell[l]{128x0e+64x1o+32x2e+32x3o} \\
\arrayrulecolor{white}\midrule
Hidden (L2) & \makecell[l]{128x0e+64x1o+64x1e\\+32x2+32x2e+32x3o+32x3e} \\
\arrayrulecolor{white}\midrule
Hidden (L3) & \makecell[l]{128x0e+128x0e+64x1o+64x1e\\+32x2o+32x2e+32x3o+32x3e} \\
\arrayrulecolor{black}
\bottomrule
\end{tabular*}
\end{subtable}%
\hspace{0.040\textwidth}%
\begin{subtable}[t]{0.460\textwidth}
\centering
\caption{BAM (40.15M)}
\begin{tabular*}{\linewidth}{@{\extracolsep{\fill}}>{\raggedright\arraybackslash}p{0.43\linewidth}|>{\raggedright\arraybackslash}p{0.52\linewidth}@{}}
\toprule
\multicolumn{1}{c|}{\textbf{Hyper-parameter}} & \multicolumn{1}{c}{\textbf{Value}} \\
\midrule
Layers & 5 \\
Cutoff radius & 5.0$~\text{\AA}$ \\
\midrule
\multicolumn{1}{c|}{\textbf{Stage}} & \multicolumn{1}{c}{\textbf{Irreps}} \\
\midrule
Edge & \makecell[l]{1x0e+1x1o\\+1x2e+1x3o} \\\\
\arrayrulecolor{white}\midrule
Input & 128x0e \\
\arrayrulecolor{white}\midrule
Hidden (L1--5) & \makecell[l]{64x0o+64x0e\\+64x1o+64x1e\\+64x2o+64x2e\\+64x3o+64x3e} \\
\arrayrulecolor{black}
\bottomrule
\end{tabular*}
\end{subtable}%

\vspace{0.6em}
\begin{minipage}{0.97\textwidth}
\scriptsize
\end{minipage}
\end{table}

\clearpage

\section{Detailed Benchmark Data}\label{app: detailed benchmark data}
\subsection{Raw Data for Kernel Microbenchmarks}\label{app: raw data for kernel bench}
%我们遵循Section~\ref{APP: evaluation setup}的实验与测试配置，对LunaTP进行性能评估。Table~\ref{tab:raw-kernel-all} reports raw data of kernel microbenchmark. 
We evaluate LunaTP using the experimental setup described in Section~\ref{app: evaluation setup}. Table~\ref{tab:raw-kernel-all} reports the raw results of our kernel microbenchmarks.

\begin{table}[h]
\centering
\caption{Kernel measurements from forward through third-order differentiation. Parentheses report runtime speedup and peak-memory reduction relative to e3nn.}
\label{tab:raw-kernel-all}
\label{tab:raw-kernel-first}
\label{tab:raw-kernel-higher}
\fontsize{8}{8.3}\selectfont
\setlength{\tabcolsep}{1.2pt}
\renewcommand{\arraystretch}{1.07}
\begin{adjustbox}{max width=\textwidth}
\begin{tabular}{@{}lccccc@{\hspace{5pt}}ccccc@{}}

\toprule
\multirow{2}{*}{Case ID} & \multicolumn{5}{c}{\textbf{Forward}} & \multicolumn{5}{c}{\textbf{Backward}} \\
\cmidrule(lr){2-6}\cmidrule(lr){7-11}
 & e3nn & FlashTP & OpenEq & cuEq & LunaTP (ours) & e3nn & FlashTP & OpenEq & cuEq & LunaTP (ours) \\
\midrule
\multicolumn{11}{c}{\textit{Runtime (ms)}} \\
\midrule
C1 & 3.56\,(1.00x) & 0.563\,(6.32x) & 0.248\,(14.4x) & 4.40\,(0.809x) & 0.0763\,(46.7x) & 9.21\,(1.00x) & 1.29\,(7.16x) & 0.331\,(27.8x) & 9.59\,(0.960x) & 0.214\,(43.1x) \\
C2 & 9.85\,(1.00x) & 1.59\,(6.19x) & 0.681\,(14.5x) & 1.12\,(8.81x) & 0.163\,(60.6x) & 24.0\,(1.00x) & 3.28\,(7.32x) & 1.21\,(19.8x) & 0.514\,(46.7x) & 0.366\,(65.6x) \\
C3 & 5.77\,(1.00x) & 0.947\,(6.09x) & 0.439\,(13.1x) & 6.80\,(0.849x) & 0.127\,(45.4x) & 16.3\,(1.00x) & 1.95\,(8.35x) & 0.610\,(26.8x) & 17.3\,(0.944x) & 0.320\,(50.9x) \\
C4 & 28.3\,(1.00x) & 4.70\,(6.03x) & 1.99\,(14.3x) & 2.16\,(13.1x) & 0.449\,(63.2x) & 76.9\,(1.00x) & 9.59\,(8.03x) & 9.40\,(8.18x) & 1.56\,(49.3x) & 1.02\,(75.8x) \\
C5 & 19.8\,(1.00x) & 3.10\,(6.39x) & 1.36\,(14.6x) & 14.7\,(1.35x) & 0.344\,(57.6x) & 67.3\,(1.00x) & 7.08\,(9.51x) & 6.40\,(10.5x) & 64.6\,(1.04x) & 1.05\,(63.8x) \\
C6 & 39.5\,(1.00x) & 6.13\,(6.45x) & 2.46\,(16.1x) & 2.93\,(13.5x) & 0.586\,(67.4x) & 130\,(1.00x) & 13.4\,(9.74x) & 13.0\,(10.0x) & 2.02\,(64.5x) & 1.53\,(85.0x) \\
C7 & 26.0\,(1.00x) & 4.33\,(6.00x) & 1.54\,(16.9x) & 17.0\,(1.53x) & 0.435\,(59.7x) & 101\,(1.00x) & 9.76\,(10.4x) & 9.53\,(10.6x) & 92.2\,(1.10x) & 1.38\,(73.3x) \\
C8 & 34.7\,(1.00x) & 5.90\,(5.88x) & 2.23\,(15.5x) & 3.84\,(9.05x) & 0.544\,(63.8x) & 133\,(1.00x) & 13.1\,(10.2x) & 11.8\,(11.3x) & 2.37\,(56.1x) & 1.54\,(86.2x) \\
C9 & 46.8\,(1.00x) & 8.67\,(5.40x) & 2.85\,(16.4x) & 27.3\,(1.71x) & 0.755\,(62.0x) & 212\,(1.00x) & 21.3\,(9.99x) & 18.6\,(11.4x) & 181\,(1.17x) & 2.79\,(76.3x) \\
C10 & 44.5\,(1.00x) & 8.42\,(5.29x) & 3.05\,(14.6x) & 6.97\,(6.39x) & 0.713\,(62.4x) & 199\,(1.00x) & 20.2\,(9.87x) & 17.0\,(11.7x) & 5.14\,(38.8x) & 2.28\,(87.4x) \\
\midrule
\multicolumn{11}{c}{\textit{Peak memory (GB)}} \\
\midrule
C1 & 1.50\,(1.00x) & 0.157\,(9.60x) & 0.156\,(9.62x) & 1.51\,(0.997x) & 0.157\,(9.59x) & 2.41\,(1.00x) & 0.331\,(7.27x) & 0.312\,(7.72x) & 1.52\,(1.59x) & 0.313\,(7.70x) \\
C2 & 4.19\,(1.00x) & 0.359\,(11.7x) & 0.359\,(11.7x) & 0.391\,(10.7x) & 0.360\,(11.7x) & 7.38\,(1.00x) & 0.775\,(9.53x) & 0.718\,(10.3x) & 0.749\,(9.85x) & 0.718\,(10.3x) \\
C3 & 2.50\,(1.00x) & 0.218\,(11.5x) & 0.218\,(11.5x) & 2.47\,(1.01x) & 0.218\,(11.5x) & 3.76\,(1.00x) & 0.471\,(7.99x) & 0.435\,(8.65x) & 2.31\,(1.63x) & 0.436\,(8.64x) \\
C4 & 12.0\,(1.00x) & 0.972\,(12.3x) & 0.972\,(12.3x) & 1.06\,(11.3x) & 0.972\,(12.3x) & 20.2\,(1.00x) & 2.10\,(9.63x) & 1.94\,(10.4x) & 2.03\,(9.93x) & 1.94\,(10.4x) \\
C5 & 8.15\,(1.00x) & 0.607\,(13.4x) & 0.606\,(13.5x) & 8.00\,(1.02x) & 0.611\,(13.4x) & 13.2\,(1.00x) & 1.33\,(9.87x) & 1.21\,(10.9x) & 7.00\,(1.88x) & 1.22\,(10.8x) \\
C6 & 14.3\,(1.00x) & 0.907\,(15.8x) & 0.907\,(15.8x) & 1.01\,(14.2x) & 0.907\,(15.8x) & 28.0\,(1.00x) & 1.98\,(14.2x) & 1.81\,(15.5x) & 1.91\,(14.6x) & 1.81\,(15.4x) \\
C7 & 9.09\,(1.00x) & 0.535\,(17.0x) & 0.535\,(17.0x) & 8.86\,(1.03x) & 0.539\,(16.9x) & 16.4\,(1.00x) & 1.18\,(13.8x) & 1.07\,(15.3x) & 7.18\,(2.28x) & 1.07\,(15.3x) \\
C8 & 12.8\,(1.00x) & 0.628\,(20.4x) & 0.628\,(20.5x) & 0.713\,(18.0x) & 0.629\,(20.4x) & 22.5\,(1.00x) & 1.39\,(16.2x) & 1.26\,(17.9x) & 1.34\,(16.8x) & 1.26\,(17.9x) \\
C9 & 16.7\,(1.00x) & 0.781\,(21.4x) & 0.781\,(21.4x) & 15.1\,(1.11x) & 0.787\,(21.3x) & 28.5\,(1.00x) & 1.64\,(17.4x) & 1.56\,(18.3x) & 11.9\,(2.40x) & 1.57\,(18.2x) \\
C10 & 16.4\,(1.00x) & 0.658\,(25.0x) & 0.658\,(25.0x) & 0.758\,(21.7x) & 0.659\,(25.0x) & 27.4\,(1.00x) & 1.37\,(20.0x) & 1.32\,(20.8x) & 1.42\,(19.4x) & 1.32\,(20.8x) \\
\midrule

\midrule

\multirow{2}{*}{Case ID} & \multicolumn{5}{c}{\textbf{Double-bwd}} & \multicolumn{5}{c}{\textbf{Triple-bwd}} \\
\cmidrule(lr){2-6}\cmidrule(lr){7-11}
 & e3nn & FlashTP & OpenEq & cuEq & LunaTP (ours) & e3nn & FlashTP & OpenEq & cuEq & LunaTP (ours) \\
\midrule
\multicolumn{11}{c}{\textit{Runtime (ms)}} \\
\midrule
C1 & 25.7\,(1.00x) & 1.95\,(13.2x) & 1.12\,(23.0x) & 25.3\,(1.02x) & 0.403\,(63.9x) & 47.2\,(1.00x) & -- & 6.38\,(7.41x) & 45.5\,(1.04x) & 0.683\,(69.2x) \\
C2 & 74.4\,(1.00x) & 4.71\,(15.8x) & 3.61\,(20.6x) & 1.43\,(52.1x) & 0.802\,(92.8x) & 132\,(1.00x) & -- & 19.9\,(6.64x) & 4.77\,(27.7x) & 1.74\,(75.9x) \\
C3 & 52.8\,(1.00x) & 2.78\,(19.0x) & 2.23\,(23.7x) & 52.0\,(1.01x) & 0.651\,(81.2x) & 89.5\,(1.00x) & -- & 12.0\,(7.48x) & 86.7\,(1.03x) & 1.17\,(76.2x) \\
C4 & 251\,(1.00x) & 14.1\,(17.8x) & 12.1\,(20.8x) & 5.11\,(49.2x) & 2.36\,(106x) & 429\,(1.00x) & -- & 82.1\,(5.22x) & 23.7\,(18.1x) & 5.14\,(83.3x) \\
C5 & 262\,(1.00x) & 10.9\,(24.1x) & 7.77\,(33.7x) & 245\,(1.07x) & 1.96\,(134x) & 401\,(1.00x) & -- & 53.5\,(7.49x) & 362\,(1.11x) & 3.98\,(101x) \\
C6 & 467\,(1.00x) & 21.0\,(22.2x) & 16.5\,(28.4x) & 12.9\,(36.3x) & 2.83\,(165x) & 758\,(1.00x) & -- & 109\,(6.93x) & 112\,(6.77x) & 6.39\,(119x) \\
C7 & 427\,(1.00x) & 15.9\,(26.8x) & 10.6\,(40.3x) & 392\,(1.09x) & 2.20\,(194x) & 629\,(1.00x) & -- & 73.8\,(8.53x) & 555\,(1.13x) & 5.13\,(123x) \\
C8 & 537\,(1.00x) & 21.5\,(25.0x) & 16.3\,(33.0x) & 22.9\,(23.5x) & 2.84\,(189x) & 813\,(1.00x) & -- & 104\,(7.85x) & 243\,(3.35x) & 7.10\,(114x) \\
C9 & 1010\,(1.00x) & 39.8\,(25.4x) & 22.2\,(45.5x) & 926\,(1.09x) & 4.04\,(250x) & 1630\,(1.00x) & -- & 148\,(11.0x) & 1230\,(1.33x) & 11.4\,(143x) \\
C10 & 895\,(1.00x) & 38.4\,(23.3x) & 22.6\,(39.5x) & 57.5\,(15.6x) & 3.70\,(242x) & 1420\,(1.00x) & -- & 145\,(9.80x) & OOM & 10.7\,(132x) \\
\midrule
\multicolumn{11}{c}{\textit{Peak memory (GB)}} \\
\midrule
C1 & 3.80\,(1.00x) & 0.614\,(6.19x) & 0.604\,(6.29x) & 3.92\,(0.969x) & 0.614\,(6.18x) & 6.33\,(1.00x) & -- & 1.98\,(3.20x) & 5.27\,(1.20x) & 0.947\,(6.69x) \\
C2 & 10.2\,(1.00x) & 1.41\,(7.25x) & 1.38\,(7.41x) & 1.44\,(7.10x) & 1.39\,(7.35x) & 19.0\,(1.00x) & -- & 4.52\,(4.20x) & 2.16\,(8.79x) & 2.18\,(8.71x) \\
C3 & 5.71\,(1.00x) & 0.855\,(6.68x) & 0.836\,(6.84x) & 6.35\,(0.900x) & 0.855\,(6.68x) & 9.94\,(1.00x) & -- & 2.74\,(3.63x) & 8.24\,(1.21x) & 1.33\,(7.49x) \\
C4 & 30.7\,(1.00x) & 3.81\,(8.07x) & 3.73\,(8.24x) & 3.90\,(7.88x) & 3.75\,(8.20x) & 54.8\,(1.00x) & -- & 12.2\,(4.48x) & 5.87\,(9.34x) & 5.93\,(9.24x) \\
C5 & 20.7\,(1.00x) & 2.37\,(8.75x) & 2.31\,(8.99x) & 20.2\,(1.03x) & 2.38\,(8.72x) & 34.9\,(1.00x) & -- & 7.58\,(4.60x) & 25.0\,(1.39x) & 3.72\,(9.38x) \\
C6 & 47.1\,(1.00x) & 3.52\,(13.4x) & 3.44\,(13.7x) & 3.64\,(13.0x) & 3.52\,(13.4x) & 75.6\,(1.00x) & -- & 11.3\,(6.68x) & 5.47\,(13.8x) & 5.56\,(13.6x) \\
C7 & 28.9\,(1.00x) & 2.07\,(14.0x) & 2.01\,(14.3x) & 22.8\,(1.27x) & 2.08\,(13.9x) & 45.6\,(1.00x) & -- & 6.64\,(6.87x) & 29.0\,(1.57x) & 3.28\,(13.9x) \\
C8 & 41.1\,(1.00x) & 2.41\,(17.1x) & 2.35\,(17.5x) & 2.52\,(16.3x) & 2.42\,(17.0x) & 63.9\,(1.00x) & -- & 7.77\,(8.23x) & 3.79\,(16.9x) & 3.84\,(16.7x) \\
C9 & 52.5\,(1.00x) & 2.99\,(17.5x) & 2.91\,(18.0x) & 38.9\,(1.35x) & 3.02\,(17.4x) & 79.8\,(1.00x) & -- & 9.39\,(8.50x) & 48.3\,(1.65x) & 4.67\,(17.1x) \\
C10 & 52.3\,(1.00x) & 2.50\,(20.9x) & 2.44\,(21.4x) & 2.64\,(19.8x) & 2.51\,(20.8x) & 79.2\,(1.00x) & -- & 7.97\,(9.93x) & OOM & 3.97\,(20.0x) \\
\bottomrule

\end{tabular}
\end{adjustbox}

\vspace{0.6em}
\begin{minipage}{1.00\textwidth}
\footnotesize 
``--'' denotes an unsupported setting.
\end{minipage}
\end{table}

\clearpage

\subsection{Raw Data for Training Benchmarks}\label{app: raw data for training}
Tables~\ref{tab:raw-training-efs} and~\ref{tab:raw-training-efsh} report throughput and the CGTP convolution time fraction for all eight models under EFS and EFSH training.

\begin{table}[!htbp]
\centering
\caption{EFS training performance. Each entry reports throughput in ``atoms/s'', with the CGTP convolution time fraction in parentheses.}
\label{tab:raw-training-efs}
\footnotesize
\setlength{\tabcolsep}{3pt}
\renewcommand{\arraystretch}{1.10}
\begin{tabular*}{\linewidth}{@{\extracolsep{\fill}}llccccc@{}}
\toprule
\multirow{2}{*}{Setting} & \multirow{2}{*}{Model} & \multicolumn{5}{c}{Throughput (CGTP time fraction)} \\
\cmidrule(lr){3-7}
 & & e3nn & FlashTP & OpenEq & cuEq & LunaTP (ours) \\
\midrule
\multirow{8}{*}{Matched} & SevenNet-Omni & 936.6\,(95.9\%) & 8470.3\,(51.2\%) & 9160.7\,(47.5\%) & 358.7\,(76.5\%) & \textbf{18024.6}\,(21.3\%) \\
 & MACE-L & 4023.5\,(53.5\%) & 7084.6\,(10.7\%) & 7435.6\,(7.9\%) & 774.6\,(59.7\%) & \textbf{8455.9}\,(1.4\%) \\
 & NequIP-XL & 625.8\,(93.8\%) & 4696.5\,(57.5\%) & 5225.1\,(48.9\%) & 370.7\,(70.3\%) & \textbf{7703.4}\,(20.2\%) \\
 & NequIX & 4929.0\,(92.8\%) & 26052.0\,(46.1\%) & 40017.6\,(41.0\%) & 4989.0\,(88.9\%) & \textbf{52397.8}\,(20.9\%) \\
 & EquFlashV2 & 144.0\,(78.4\%) & 547.1\,(26.1\%) & 538.0\,(19.8\%) & 504.6\,(9.5\%) & \textbf{659.7}\,(6.0\%) \\
 & TACE-L & 477.8\,(79.0\%) & 1625.5\,(28.6\%) & 1748.3\,(23.3\%) & 245.5\,(69.5\%) & \textbf{2099.9}\,(4.6\%) \\
 & Prophet & 160.2\,(84.3\%) & 724.5\,(31.5\%) & 782.7\,(24.8\%) & 149.2\,(50.3\%) & \textbf{824.7}\,(11.0\%) \\
 & BAM & 531.4\,(82.3\%) & 2654.6\,(16.3\%) & 2538.7\,(15.7\%) & 601.7\,(80.0\%) & \textbf{2824.5}\,(3.1\%) \\
\midrule
\multirow{8}{*}{MaxBS} & SevenNet-Omni & 936.6\,(95.1\%) & 11136.3\,(71.6\%) & 12469.2\,(61.6\%) & 361.1\,(82.8\%) & \textbf{23933.1}\,(26.1\%) \\
 & MACE-L & 4023.5\,(53.5\%) & 7521.5\,(11.3\%) & 7759.8\,(8.6\%) & 1019.9\,(78.0\%) & \textbf{9289.7}\,(1.5\%) \\
 & NequIP-XL & 625.8\,(93.8\%) & 5648.9\,(69.1\%) & 6961.3\,(64.0\%) & 368.8\,(71.1\%) & \textbf{12844.3}\,(26.2\%) \\
 & NequIX & 4929.0\,(92.8\%) & 24715.1\,(43.4\%) & 42825.2\,(44.1\%) & 5046.9\,(66.8\%) & \textbf{52424.9}\,(20.9\%) \\
 & EquFlashV2 & 144.0\,(78.4\%) & 1390.4\,(65.6\%) & 1577.2\,(60.5\%) & 2147.8\,(38.1\%) & \textbf{3311.9}\,(19.5\%) \\
 & TACE-L & 477.8\,(79.0\%) & 2614.6\,(46.6\%) & 2858.3\,(40.5\%) & 242.6\,(69.5\%) & \textbf{4563.7}\,(9.1\%) \\
 & Prophet & 160.2\,(84.3\%) & 1076.0\,(43.9\%) & 1174.9\,(38.6\%) & 169.0\,(53.0\%) & \textbf{1583.3}\,(14.2\%) \\
 & BAM & 531.4\,(82.3\%) & 2659.0\,(15.7\%) & 2640.0\,(15.9\%) & 618.5\,(80.0\%) & \textbf{2998.9}\,(3.2\%) \\
\bottomrule
\end{tabular*}
\end{table}

\begin{table}[!htbp]
\centering
\caption{EFSH training performance. Each entry reports throughput in ``atoms/s'', with the tensor product convolution time fraction in parentheses.}
\label{tab:raw-training-efsh}
\footnotesize
\setlength{\tabcolsep}{3pt}
\renewcommand{\arraystretch}{1.10}
\begin{tabular*}{\linewidth}{@{\extracolsep{\fill}}llccccc@{}}
\toprule
\multirow{2}{*}{Setting} & \multirow{2}{*}{Model} & \multicolumn{5}{c}{Throughput (CGTP time fraction)} \\
\cmidrule(lr){3-7}
 & & e3nn & FlashTP & OpenEq & cuEq & LunaTP (ours) \\
\midrule
\multirow{8}{*}{Matched} & SevenNet-Omni & 277.5\,(93.4\%) & -- & 2036.4\,(52.4\%) & 156.1\,(71.7\%) & \textbf{3232.3}\,(18.0\%) \\
 & MACE-L & 1204.6\,(56.1\%) & -- & 2371.2\,(13.5\%) & 486.3\,(48.7\%) & \textbf{2775.5}\,(2.3\%) \\
 & NequIP-XL & 182.5\,(92.0\%) & -- & 1100.4\,(52.9\%) & -- & \textbf{1480.7}\,(22.6\%) \\
 & NequIX & 1465.7\,(89.9\%) & -- & 8352.2\,(56.3\%) & 1506.9\,(91.8\%) & \textbf{14369.5}\,(25.0\%) \\
 & EquFlashV2 & 47.0\,(55.4\%) & -- & 85.6\,(11.2\%) & 72.5\,(11.6\%) & \textbf{91.4}\,(9.2\%) \\
 & TACE-L & 122.6\,(47.0\%) & -- & 202.5\,(12.5\%) & 116.3\,(52.2\%) & \textbf{221.5}\,(3.5\%) \\
 & Prophet & 34.0\,(53.9\%) & -- & 76.0\,(6.3\%) & 29.8\,(64.7\%) & \textbf{80.7}\,(4.2\%) \\
 & BAM & 182.9\,(83.7\%) & -- & 949.9\,(26.7\%) & 179.7\,(86.1\%) & \textbf{1138.2}\,(6.2\%) \\
\midrule
\multirow{8}{*}{MaxBS} & SevenNet-Omni & 277.5\,(93.4\%) & -- & 2562.0\,(72.2\%) & 166.9\,(77.8\%) & \textbf{7956.0}\,(30.2\%) \\
 & MACE-L & 1204.6\,(56.1\%) & -- & 2979.7\,(9.5\%) & 755.92\,(87.7\%) & \textbf{4248.1}\,(2.7\%) \\
 & NequIP-XL & 182.5\,(92.0\%) & -- & 1490.5\,(78.0\%) & -- & \textbf{4627.4}\,(33.4\%) \\
 & NequIX & 1465.7\,(89.9\%) & -- & 9082.6\,(60.0\%) & 1401.0\,(92.9\%) & \textbf{16450.9}\,(27.0\%) \\
 & EquFlashV2 & 42.7\,(55.4\%) & -- & 307.1\,(64.3\%) & 247.5\,(64.1\%) & \textbf{704.6}\,(17.6\%) \\
 & TACE-L & 122.6\,(46.8\%) & -- & 635.7\,(50.6\%) & 150.3\,(69.3\%) & \textbf{1233.1}\,(9.8\%) \\
 & Prophet & 39.1\,(89.3\%) & -- & 256.0\,(43.3\%) & 36.0\,(64.4\%) & \textbf{408.5}\,(15.8\%) \\
 & BAM & 182.9\,(83.7\%) & -- & 1029.6\,(30.1\%) & 179.7\,(86.1\%) & \textbf{1386.7}\,(5.4\%) \\
\bottomrule
\end{tabular*}

\vspace{0.6em}
\begin{minipage}{\linewidth}
\footnotesize
``--'' denotes an unsupported setting.
\end{minipage}
\end{table}

\clearpage

\section{Ablation Studies}\label{sec: ablation}
We evaluate the effects of path fusion scope, cross-path input reuse, task ordering, and compile-time reuse on LunaTP's forward performance. All experiments use the C5 configuration in Table~\ref{tab:kernel-workloads} with FP32 precision. This configuration includes channel counts of 320, 96, 64, and 32.

\subsection{Path Grouping}\label{sec: path grouping}
Path grouping needs to balance input reuse and accumulator storage. We first vary the per-channel output component budget $\beta$ to examine the effect of fusion scope (Table~\ref{tab:ablation-scope}). Among the tested settings, the default budget $\beta=16$ gives the lowest latency. Fusing all compatible paths increases latency by 56.3\%, while reducing the budget to 8 and 4 increases latency by 11.6\% and 20.8\%, respectively. Smaller groups require more repeated edge traversals, while full fusion keeps more accumulators live at once and is the only tested setting with reported register spilling.

Since the default policy automatically adjusts the number of warps as the grouping changes, we also compare the fusion scopes with fixed warp counts. With 8 and 16 warps, full fusion is still 44.3\% (0.5001 vs.\ 0.3467\,ms) and 53.0\% (0.5423 vs.\ 0.3545\,ms) slower, respectively. In both cases, the reported spill stores are 2976\,B for full fusion and zero for the default budget.

\begin{table}[h]
  \centering
  \caption{Output component budget with automatic warp selection (C5, FP32, forward). Latency is the median of two independent measurements. Spill stores are kernel-reported statistics, not total runtime traffic.}
  \label{tab:ablation-scope}
  \begin{tabular}{lrrrrr}
    \toprule
    Configuration & Path groups & Execution units & Registers & Spill stores (B) & Latency (ms) \\
    \midrule
    All compatible paths & 5 & 17 & 64 & 2976 & 0.5393 \\
    Default ($\beta=16$) & 17 & 45 & 64 & 0 & 0.3449 \\
    $\beta=8$ & 29 & 73 & 56 & 0 & 0.3851 \\
    $\beta=4$ & 37 & 96 & 56 & 0 & 0.4167 \\
    \bottomrule
  \end{tabular}
\end{table}

Performance also depends on which paths can share inputs within a group. Compared with the baseline that groups paths by edge irrep, broader cross-path input reuse reduces the number of path groups from 25 to 17 and lowers latency by about 6.6\% (Table~\ref{tab:ablation-comp}). This change both expands input reuse and reduces the number of tasks, so the measured gain reflects the overall effect of the grouping change. With this grouping, enabling the work partitioning option shown in the table leaves latency nearly unchanged, with no additional benefit observed on C5.

\begin{table}[h]
  \centering
  \caption{Cross-path input reuse and work partitioning (C5, FP32, forward). All configurations use node-centric execution.}
  \label{tab:ablation-comp}
  \begin{tabular}{lrrr}
    \toprule
    Configuration & Path groups & Execution units & Latency (ms) \\
    \midrule
    Baseline (grouped by edge irrep) & 25 & 85 & 0.3688 \\
    $+$ Cross-path input reuse & 17 & 45 & 0.3446 \\
    $+$ Work partitioning (full) & 17 & 45 & 0.3447 \\
    \bottomrule
  \end{tabular}
\end{table}

\subsection{Task Ordering}\label{sec: task ordering}
We next vary the task order while keeping the fusion scope and number of warps fixed. We independently reverse the ordering by nonzero CG-term count and by node degree, while keeping the task set unchanged (Table~\ref{tab:ablation-order}). Compared with descending order on both axes, changing only the node order to ascending increases latency by 13.0\%. Changing only the CG-term count order to ascending increases latency by 5.8\%, while changing both increases it by 17.1\%. Sorting both in descending order, as used in the main text, gives the lowest latency among these settings.

\begin{table}[h]
  \centering
  \caption{Task ordering with a fixed fusion scope and warp count (C5, FP32, forward). Each ordering axis is reversed independently; the task set, instruction set, and output are unchanged.}
  \label{tab:ablation-order}
  \begin{tabular}{rrrr}
    \toprule
    CG-term count order & Node degree order & Latency (ms) & vs.\ both descending \\
    \midrule
    Descending & Descending & 0.3446 & --- \\
    Descending & Ascending & 0.3895 & $+13.0\%$ \\
    Ascending & Descending & 0.3645 & $+5.8\%$ \\
    Ascending & Ascending & 0.4036 & $+17.1\%$ \\
    \bottomrule
  \end{tabular}
\end{table}

\subsection{Compile-Time Reuse}\label{sec: compile-time reuse}
Finally, we compare all four combinations of product reuse and coefficient reuse with a fixed fusion scope and number of warps (Table~\ref{tab:ablation-reuse}). Enabling both gives the lowest latency of 0.3449\,ms. Disabling coefficient reuse increases latency by 3.7\%, while disabling both increases it by 3.4\%.

These two switches are not fully independent. The code generation path for coefficient reuse retains shared products, so disabling the product reuse switch does not necessarily remove product sharing from these computations. When coefficient reuse is disabled, disabling product reuse also does not increase latency.

\begin{table}[h]
  \centering
  \caption{Compile-time reuse switches with fixed fusion scope and warp count (C5, FP32, forward). Coefficient reuse retains shared products even when the product-reuse switch is disabled.}
  \label{tab:ablation-reuse}
  \begin{tabular}{ccrr}
    \toprule
    Product reuse & Coefficient reuse & Latency (ms) & vs.\ both enabled \\
    \midrule
    Enabled & Enabled & 0.3449 & --- \\
    Enabled & Disabled & 0.3578 & $+3.7\%$ \\
    Disabled & Enabled & 0.3526 & $+2.2\%$ \\
    Disabled & Disabled & 0.3565 & $+3.4\%$ \\
    \bottomrule
  \end{tabular}
\end{table}

\section{Additional Kernel Microbenchmark Results}\label{app: addition kernel bench}
\subsection{Kernel microbenchmark (FP64)}\label{sec:kernel result fp64}
We evaluate FP64 performance using the same irrep configurations as in the FP32 benchmarks. Since FP64 requires more computation and memory, we reduce the atom count by half to 384 for most cases and further reduce it to 192 for C8 and C10. All runtime speedups and peak-memory reduction factors are relative to FP64 e3nn on the same hardware.
\begin{figure}[h]
\begin{center}
    \includegraphics[width=1.0\linewidth]{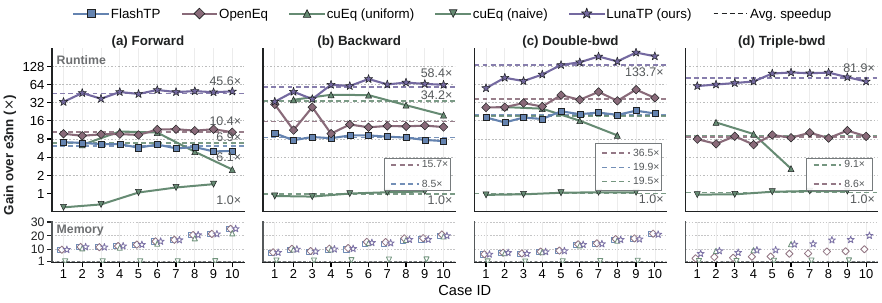}
    \captionsetup{justification=centering,singlelinecheck=false}
    \caption{ 
        LunaTP performance in kernel microbenchmarks (FP64, NVIDIA H100-80GB).
    }
    \label{fig: training performance on h100 fp64}
\end{center}
\end{figure}
%In this section, 我们在相同的 irrep 配置上进行 FP64 微基准测试。Since FP64 更高的计算与显存开销，我们将测试原子数通常减半至 384，C8 和 C10 则进一步缩减至 192。所有运行时间加速比和显存降低倍数均以 FP64 e3nn 在相同硬件上的结果为基准。
%我们采用与 FP32 测试相同的 irrep 配置评估 FP64 性能，并考虑其更大的显存需求，将每个负载设为 384 个原子，C8 和 C10 则使用 192 个原子。如图~\ref{fig: training performance on h100 fp64} 所示，LunaTP 相对 FP64 e3nn 在前向、反向、二阶反向和三阶反向中的算术平均加速比分别为 $45.6\times$、$58.4\times$、$133.7\times$ 和 $81.9\times$；相对 OpenEq 则分别为 $4.42\times$、$4.40\times$、$3.56\times$ 和 $9.61\times$。
Figure~\ref{fig: training performance on h100 fp64} summarizes the results. LunaTP achieves arithmetic-mean speedups of 45.6$\times$, 58.4$\times$, 133.7$\times$, and 81.9$\times$ over FP64 e3nn for forward, backward, double backward, and triple backward, respectively. The corresponding speedups over OpenEq are 4.42$\times$, 4.40$\times$, 3.56$\times$, and 9.61$\times$.

\subsection{Kernel microbenchmark (A100 \& RTX 3090)}\label{sec:kernel result portability}
%为评估 LunaTP 在不同 GPU 上的性能，我们在 NVIDIA A100-80GB 和 RTX 3090-24GB 上重复 FP32 微基准测试。A100 沿用正文的工作负载。受显存容量限制，RTX 3090 默认使用 192 个原子，C8 和 C10 分别缩减至 122 和 97 个原子。同一 case 的所有后端和求导阶段均使用相同的负载。
We evaluate LunaTP on NVIDIA A100-80GB and RTX 3090-24GB GPUs using FP32 microbenchmarks. On A100, we use the same workloads as in the main evaluation. To fit within the RTX 3090's memory capacity, we use 192 atoms by default and reduce C8 and C10 to 122 and 97 atoms, respectively. For each case, we keep the workload fixed across all backends and all four stages.

\begin{figure}[h]
\begin{center}
    \includegraphics[width=1.0\linewidth]{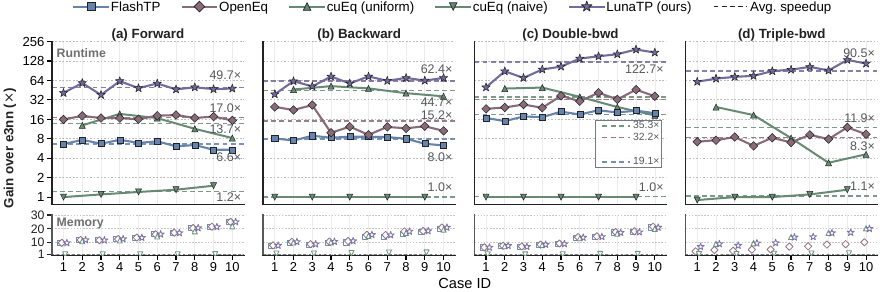}
    \captionsetup{justification=centering,singlelinecheck=false}
    \caption{ 
        LunaTP performance in kernel microbenchmarks (NVIDIA A100-80GB PCIe).
    }
    \label{fig: training performance on A100}
\end{center}
\end{figure}

\begin{figure}[h]
\begin{center}
    \includegraphics[width=1.0\linewidth]{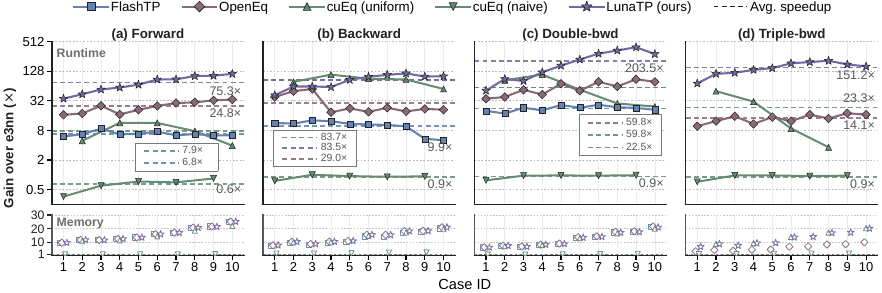}
    \captionsetup{justification=centering,singlelinecheck=false}
    \caption{ 
        LunaTP performance in kernel microbenchmarks (RTX 3090-24GB).
    }
    \label{fig: training performance on 3090}
\end{center}
\end{figure}

%在 A100 上，LunaTP 相对 e3nn 在前向、一阶、二阶和三阶反向计算中的平均加速比分别为 49.7、62.4、122.7 和 90.5 倍。在 RTX 3090 上，对应的加速比分别为 75.3、83.5、203.5 和 151.2 倍。完整结果分别见 Figure~\ref{fig: training performance on A100} 和 Figure~\ref{fig: training performance on 3090}。
On A100, LunaTP achieves average speedups of 49.7$\times$, 62.4$\times$, 122.7$\times$, and 90.5$\times$ over e3nn for forward, backward, double backward, and triple backward, respectively. On RTX 3090, the corresponding speedups are 75.3$\times$, 83.5$\times$, 203.5$\times$, and 151.2$\times$. Figures~\ref{fig: training performance on A100} and~\ref{fig: training performance on 3090} show the full results.

%LunaTP 在数据中心和消费级 GPU 上均具有性能优势，但加速幅度因硬件而异。在全部四个计算阶段中，其相对 e3nn 的平均加速比均在 RTX 3090 上最高，其次为 H100 和 A100。这一排序与三者的 FP32 峰值算力与显存带宽之比一致。这是因为访存是 e3nn 张量积计算的主要瓶颈~\citep{flashtp}，LunaTP 通过核函数融合和数据复用减少中间数据访存，在算力相对充裕、显存带宽相对有限的 GPU 上带来更大的加速收益。
LunaTP delivers performance gains on both data-center and consumer GPUs, although the speedups vary across hardware. Across all four stages, its average speedups over e3nn are highest on RTX 3090, followed by H100 and A100. This ordering matches that of the GPUs' peak FP32 compute-to-memory-bandwidth ratios. Since e3nn's tensor-product computations are primarily limited by memory bandwidth~\citep{lee2025flashtp}, LunaTP's reductions in intermediate memory traffic through kernel fusion and data reuse yield larger gains on GPUs with greater compute capacity relative to memory bandwidth.

\section{Details on Numerical Stability and Training Convergence}\label{app:numerical stability and training convergence}
\subsection{Operator-Level Numerical Stability}\label{app: numerical stability}
%遵循 \citet{shah2024flashattention} 的数值验证流程，我们以相同输入下的 e3nn FP64 输出为参考，计算 e3nn 和 LunaTP 在 FP32 下得到的输出及对应导数的 RMSE。数据构造遵循附录~\ref{APP: evaluation setup} 中的设置，完整结果见表~\ref{tab:numerical stability}。表中每项数值均为表~\ref{tab:kernel-workloads} 中所有 case 对应 RMSE 的平均值。
Following the numerical validation protocol of \citet{shah2024flashattention}, we use the e3nn FP64 as the reference. We run e3nn in FP32 and LunaTP in both FP32 and FP64 on the same inputs, and report RMSE for the forward and each derivative stage. The inputs are constructed as described in Appendix~\ref{app: evaluation setup}. Table~\ref{tab:numerical stability} reports the full results, averaged across all cases in Table~\ref{tab:kernel-workloads}.
\begin{table}[h]
\centering
\captionsetup{justification=centering,singlelinecheck=false}
\caption{Numerical error comparison between e3nn and LunaTP.}
\label{tab:numerical stability}
\small
\setlength{\tabcolsep}{7pt}
\renewcommand{\arraystretch}{1.08}

\begin{tabular}{llccc}
\toprule
Stage & Output & e3nn (FP32) & LunaTP (FP32) & LunaTP (FP64) \\
\midrule

Forward
& $\mathbf{x}'$ & 9.910429e-07 & 9.978108e-07 & 9.517594e-16  \\
\midrule

\multirow{3}{*}{Backward}
& $\bar{\mathbf{x}}$ & 3.582043e-06 & 6.545904e-06 & 6.437515e-15 \\
& $\bar{\mathbf{y}}$ & 3.516555e-06 & 2.934879e-06 & 8.966707e-15 \\
& $\bar{\mathbf{w}}$ & 1.696597e-07 & 1.712254e-07 & 4.281635e-16 \\
\midrule

\multirow{4}{*}{Double bwd}
& $\mathbf{h}_x$  & 5.083432e-06 & 7.955895e-06 & 1.274438e-14 \\
& $\mathbf{h}_y$  & 5.002968e-06 & 4.300956e-06 & 5.290353e-15 \\
& $\mathbf{h}_w$  & 2.546925e-07 & 2.582805e-07 & 6.404447e-16 \\
& $\mathbf{h}_g$ & 1.769849e-06 & 2.792170e-06 & 7.214677e-15 \\
\midrule

\multirow{7}{*}{Triple bwd}
& $\mathbf{t}_x$   & 5.818900e-06 & 6.511971e-06 & 9.811956e-15 \\
& $\mathbf{t}_y$   & 6.808703e-06 & 6.200321e-06 & 9.310777e-15 \\
& $\mathbf{t}_w$   & 3.152390e-07 & 2.198522e-07 & 6.591994e-16 \\
& $\mathbf{t}_g$   & 2.530354e-06 & 3.285050e-06 & 5.920629e-15 \\
& $\mathbf{t}_{v_x}$  & 5.368344e-06 & 5.646177e-06 & 8.875174e-15 \\
& $\mathbf{t}_{v_y}$  & 5.896732e-06 & 5.317266e-06 & 9.918370e-15 \\
& $\mathbf{t}_{v_w}$ & 2.677879e-07 & 2.514537e-07 & 6.436796e-16 \\
\bottomrule
\end{tabular}
\end{table}

\subsection{End-to-End Training Convergence}\label{app: training convergence}
\begin{figure}[h]
\begin{center}
    \includegraphics[width=1.0\linewidth]{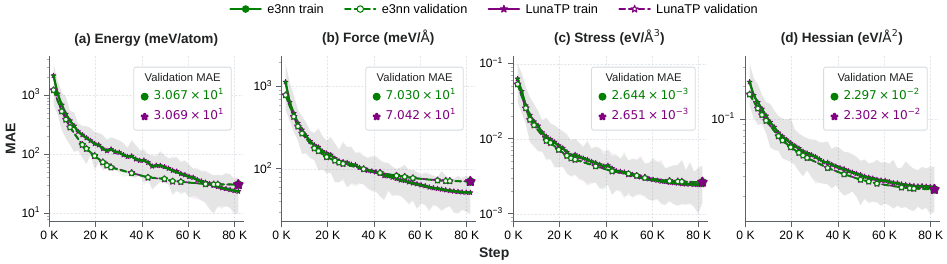}
    \captionsetup{justification=centering,singlelinecheck=false}
    \caption{ 
    Loss convergence on MPTrj-PFT dataset.
    }
    \label{fig: training convergence}
\end{center}
\end{figure}

%我们在 MPTrj-PFT 数据集上训练 NequIP-XL，以评估 LunaTP 在端到端训练中的收敛性。为保证公平比较，e3nn 与 LunaTP 使用完全相同的模型初始化、数据集和训练配置，并均采用 FP32 精度。能量、力、应力和 Hessian 的损失权重分别设为 20、20、100 和 100。模型共训练 50 个 epoch，学习率采用余弦衰减，从初始的 $5\times10^{-4}$ 降至最后一个 epoch 的 $1\times10^{-6}$。训练集与验证集的收敛曲线如 Figure~\ref{fig: training convergence} 所示。
We evaluate end-to-end training convergence by training NequIP-XL on MPTrj-PFT with either e3nn or LunaTP as the CGTP backend. Both runs use FP32 precision with the same model initialization, data, and training settings. The loss weights for energy, forces, stress, and Hessian are set to 20, 20, 100, and 100, respectively. We train for 50 epochs with a cosine learning-rate schedule, starting at $5\times10^{-4}$ and ending at $1\times10^{-6}$. Figure~\ref{fig: training convergence} shows the training and validation MAE curves for both backends.

\section{Differentiation of CGTP convolution}
\label{app: derivation of tp conv}
%在本节中，我们推导张量积卷积的一阶、二阶和三阶导数的表达式。这些表达式直接计算所需的梯度乘积，而无需具体化完整的雅可比矩阵、海森矩阵或三阶导数张量，为我们的导数核函数提供了数学基础。在微分过程中，所有导数均按公式~\ref{eq:cgtp} 以分量形式写出。
In this section, we derive the first-, second-, and third-order derivatives of tensor product convolution. The resulting formulas evaluate the gradients needed at each order without forming full Jacobian, Hessian, or third-order derivative tensors, and provide the basis for our derivative kernels. Throughout, we present the derivatives componentwise, following Eq.~\ref{eq:cgtp}.
For each channel--path combination, $\mathbf{x}_j^{(l_1,c)}\in
\mathbb{R}^{2l_1+1}$, $\mathbf{y}_{ji}^{(l_2)}\in
\mathbb{R}^{2l_2+1}$, and $w_{ji}^{(l_1,l_2,l,c)}\in\mathbb{R}$.

\paragraph{\textbf{The first-order derivative.}}
For the convolution output ${\mathbf{x}'}_i^{(l,c)}$, let
$\mathbf{g}_i^{(l,c)}=\partial\mathcal{L}/\partial{\mathbf{x}'}_i^{(l,c)}$
denote its gradient. Applying the chain rule to the weighted CGTP gives the first-order derivatives in Algorithm~\ref{alg:tpconv_first}.

\begin{algorithm}[h]
\caption{First-Order Derivative of Clebsch--Gordan Tensor Product Convolution}
\label{alg:tpconv_first}
\begin{algorithmic}[1]
\Statex \textbf{Input:} $\mathbf{x}_j^{(l_1,c)}$: node irreps
\Statex \textbf{Input:} $\mathbf{y}_{ji}^{(l_2)}$: angular edge irreps
\Statex \textbf{Input:} $w_{ji}^{(l_1,l_2,l,c)}$: path weights
\Statex \textbf{Input:} $\mathbf{g}_i^{(l,c)}$: gradients with respect to the convolution output
\Statex \textbf{Input:} edges $ji$, from source node $j$ to destination node $i$
\Statex \textbf{Output:} $\bar{\mathbf{x}}$, $\bar{\mathbf{y}}$, and $\bar{\mathbf{w}}$
\State Initialize all three outputs to zero
\For{$ji$ in the $E$ edges}
    \For{$c$ in the $C$ channels}
        \For{$(l_1,l_2)\to l$ in the $P$ paths}

            \State $\displaystyle
            \begin{aligned}
            \bar{x}_{j,m_1}^{(l_1,c)}
            \mathrel{+}={}&
            w_{ji}^{(l_1,l_2,l,c)}
            \sum_{m=-l}^{l}
            \sum_{m_2=-l_2}^{l_2}
            C^{(l,m)}_{(l_1,m_1)(l_2,m_2)}
            y_{ji,m_2}^{(l_2)}
            g_{i,m}^{(l,c)}
            \end{aligned}$

            \State $\displaystyle
            \begin{aligned}
            \bar{y}_{ji,m_2}^{(l_2)}
            \mathrel{+}={}&
            w_{ji}^{(l_1,l_2,l,c)}
            \sum_{m=-l}^{l}
            \sum_{m_1=-l_1}^{l_1}
            C^{(l,m)}_{(l_1,m_1)(l_2,m_2)}
            x_{j,m_1}^{(l_1,c)}
            g_{i,m}^{(l,c)}
            \end{aligned}$

            \State $\displaystyle
            \bar{w}_{ji}^{(l_1,l_2,l,c)}
            \mathrel{+}=
            \sum_{m=-l}^{l}
            g_{i,m}^{(l,c)}
            \left[
                \mathbf{x}_j^{(l_1,c)}
                \otimes
                \mathbf{y}_{ji}^{(l_2)}
            \right]_{m}^{(l)}$

        \EndFor
    \EndFor
\EndFor
\end{algorithmic}
\end{algorithm}

\clearpage
\paragraph{\textbf{The second-order derivative.}}
Let $\mathbf{v}_x$, $\mathbf{v}_y$, and $v_w$ denote the gradients with respect to $\bar{\mathbf{x}}$, $\bar{\mathbf{y}}$, and $\bar{w}$, respectively. Applying the product rule to the first-order formulas gives the second-order derivatives in Algorithm~\ref{alg:tpconv_second}.

\begin{algorithm}[h]
\caption{Second-Order Derivative of Clebsch--Gordan Tensor Product Convolution}
\label{alg:tpconv_second}
\begin{algorithmic}[1]
\Statex \textbf{Input:} $\mathbf{x}_j^{(l_1,c)}$: node irreps
\Statex \textbf{Input:} $\mathbf{y}_{ji}^{(l_2)}$: angular edge irreps
\Statex \textbf{Input:} $w_{ji}^{(l_1,l_2,l,c)}$: path weights
\Statex \textbf{Input:} $\mathbf{g}_i^{(l,c)}$: gradients with respect to the convolution output
\Statex \textbf{Input:} $\mathbf{v}_{x,j}^{(l_1,c)}$,
$\mathbf{v}_{y,ji}^{(l_2)}$, and
$v_{w,ji}^{(l_1,l_2,l,c)}$: gradients with respect to the first-order outputs
\Statex \textbf{Input:} edges $ji$, from source node $j$ to destination node $i$
\Statex \textbf{Output:} $\mathbf{h}_x$, $\mathbf{h}_y$,
$\mathbf{h}_w$, and $\mathbf{h}_g$
\State Initialize all four outputs to zero
\For{$ji$ in the $E$ edges}
    \For{$c$ in the $C$ channels}
        \For{$(l_1,l_2)\to l$ in the $P$ paths}

            \State $\displaystyle
            \begin{aligned}
            h_{x,j,m_1}^{(l_1,c)}
            \mathrel{+}={}&
            \sum_{m=-l}^{l}
            \sum_{m_2=-l_2}^{l_2}
            C^{(l,m)}_{(l_1,m_1)(l_2,m_2)}
            g_{i,m}^{(l,c)}
            \Bigl(
                v_{y,ji,m_2}^{(l_2)}
                w_{ji}^{(l_1,l_2,l,c)}
                +
                y_{ji,m_2}^{(l_2)}
                v_{w,ji}^{(l_1,l_2,l,c)}
            \Bigr)
            \end{aligned}$

            \State $\displaystyle
            \begin{aligned}
            h_{y,ji,m_2}^{(l_2)}
            \mathrel{+}={}&
            \sum_{m=-l}^{l}
            \sum_{m_1=-l_1}^{l_1}
            C^{(l,m)}_{(l_1,m_1)(l_2,m_2)}
            g_{i,m}^{(l,c)}
            \Bigl(
                v_{x,j,m_1}^{(l_1,c)}
                w_{ji}^{(l_1,l_2,l,c)}
                +
                x_{j,m_1}^{(l_1,c)}
                v_{w,ji}^{(l_1,l_2,l,c)}
            \Bigr)
            \end{aligned}$

            \State $\displaystyle
            \begin{aligned}
            h_{w,ji}^{(l_1,l_2,l,c)}
            \mathrel{+}={}&
            \sum_{m=-l}^{l}
            g_{i,m}^{(l,c)}
            \Biggl(
                \left[
                    \mathbf{v}_{x,j}^{(l_1,c)}
                    \otimes
                    \mathbf{y}_{ji}^{(l_2)}
                \right]_{m}^{(l)}
                +
                \left[
                    \mathbf{x}_j^{(l_1,c)}
                    \otimes
                    \mathbf{v}_{y,ji}^{(l_2)}
                \right]_{m}^{(l)}
            \Biggr)
            \end{aligned}$

            \State $\displaystyle
            \begin{aligned}
            \mathbf{h}_{g,i}^{(l,c)}
            \mathrel{+}={}&
            w_{ji}^{(l_1,l_2,l,c)}
            \Biggl(
                \left[
                    \mathbf{v}_{x,j}^{(l_1,c)}
                    \otimes
                    \mathbf{y}_{ji}^{(l_2)}
                \right]^{(l)}
                +
                \left[
                    \mathbf{x}_j^{(l_1,c)}
                    \otimes
                    \mathbf{v}_{y,ji}^{(l_2)}
                \right]^{(l)}
            \Biggr)
            \\
            &+
            v_{w,ji}^{(l_1,l_2,l,c)}
            \left[
                \mathbf{x}_j^{(l_1,c)}
                \otimes
                \mathbf{y}_{ji}^{(l_2)}
            \right]^{(l)}
            \end{aligned}$

        \EndFor
    \EndFor
\EndFor
\end{algorithmic}
\end{algorithm}

\paragraph{\textbf{The third-order derivative.}}
Let $\mathbf{q}_x$, $\mathbf{q}_y$, $q_w$, and $\mathbf{q}_g$ denote the gradients with respect to $\mathbf{h}_x$, $\mathbf{h}_y$, $\mathbf{h}_w$, and $\mathbf{h}_g$, respectively. Differentiating the second-order formulas gives the third-order derivatives in Algorithm~\ref{alg:tpconv_third}.

\begin{algorithm}[h]
\caption{Third-Order Derivative of Clebsch--Gordan Tensor Product Convolution}
\label{alg:tpconv_third}
\begin{algorithmic}[1]
\Statex \textbf{Input:} $\mathbf{x}_j^{(l_1,c)}$: node irreps
\Statex \textbf{Input:} $\mathbf{y}_{ji}^{(l_2)}$: angular edge irreps
\Statex \textbf{Input:} $w_{ji}^{(l_1,l_2,l,c)}$: path weights
\Statex \textbf{Input:} $\mathbf{g}_i^{(l,c)}$: gradients with respect to the convolution output
\Statex \textbf{Input:} $\mathbf{v}_{x,j}^{(l_1,c)}$,
$\mathbf{v}_{y,ji}^{(l_2)}$, and
$v_{w,ji}^{(l_1,l_2,l,c)}$: gradients with respect to the first-order outputs
\Statex \textbf{Input:} $\mathbf{q}_{x,j}^{(l_1,c)}$,
$\mathbf{q}_{y,ji}^{(l_2)}$,
$q_{w,ji}^{(l_1,l_2,l,c)}$, and
$\mathbf{q}_{g,i}^{(l,c)}$: gradients with respect to the second-order outputs
\Statex \textbf{Input:} edges $ji$, from source node $j$ to destination node $i$
\Statex \textbf{Output:} $\mathbf{t}_x$, $\mathbf{t}_y$,
$\mathbf{t}_w$, $\mathbf{t}_g$, $\mathbf{t}_{v_x}$,
$\mathbf{t}_{v_y}$, and $\mathbf{t}_{v_w}$
\State Initialize all seven outputs to zero
\For{$ji$ in the $E$ edges}
    \For{$c$ in the $C$ channels}
        \For{$(l_1,l_2)\to l$ in the $P$ paths}

            \State $\displaystyle
            \begin{aligned}
            t_{x,j,m_1}^{(l_1,c)}
            \mathrel{+}={}&
            \sum_{m=-l}^{l}
            \sum_{m_2=-l_2}^{l_2}
            C^{(l,m)}_{(l_1,m_1)(l_2,m_2)}
            \Bigl[
                g_{i,m}^{(l,c)}
                \Bigl(
                    q_{y,ji,m_2}^{(l_2)}
                    v_{w,ji}^{(l_1,l_2,l,c)}
                    +
                    v_{y,ji,m_2}^{(l_2)}
                    q_{w,ji}^{(l_1,l_2,l,c)}
                \Bigr)
            \\
            &\qquad+
                q_{g,i,m}^{(l,c)}
                \Bigl(
                    v_{y,ji,m_2}^{(l_2)}
                    w_{ji}^{(l_1,l_2,l,c)}
                    +
                    y_{ji,m_2}^{(l_2)}
                    v_{w,ji}^{(l_1,l_2,l,c)}
                \Bigr)
            \Bigr]
            \end{aligned}$

            \State $\displaystyle
            \begin{aligned}
            t_{y,ji,m_2}^{(l_2)}
            \mathrel{+}={}&
            \sum_{m=-l}^{l}
            \sum_{m_1=-l_1}^{l_1}
            C^{(l,m)}_{(l_1,m_1)(l_2,m_2)}
            \Bigl[
                g_{i,m}^{(l,c)}
                \Bigl(
                    q_{x,j,m_1}^{(l_1,c)}
                    v_{w,ji}^{(l_1,l_2,l,c)}
                    +
                    v_{x,j,m_1}^{(l_1,c)}
                    q_{w,ji}^{(l_1,l_2,l,c)}
                \Bigr)
            \\
            &\qquad+
                q_{g,i,m}^{(l,c)}
                \Bigl(
                    v_{x,j,m_1}^{(l_1,c)}
                    w_{ji}^{(l_1,l_2,l,c)}
                    +
                    x_{j,m_1}^{(l_1,c)}
                    v_{w,ji}^{(l_1,l_2,l,c)}
                \Bigr)
            \Bigr]
            \end{aligned}$

            \State $\displaystyle
            \begin{aligned}
            t_{w,ji}^{(l_1,l_2,l,c)}
            \mathrel{+}={}&
            \sum_{m=-l}^{l}
            \Biggl(
                g_{i,m}^{(l,c)}
                \Biggl(
                    \left[
                        \mathbf{q}_{x,j}^{(l_1,c)}
                        \otimes
                        \mathbf{v}_{y,ji}^{(l_2)}
                    \right]_{m}^{(l)}
                    +
                    \left[
                        \mathbf{v}_{x,j}^{(l_1,c)}
                        \otimes
                        \mathbf{q}_{y,ji}^{(l_2)}
                    \right]_{m}^{(l)}
                \Biggr)
            \\
            &\qquad+
                q_{g,i,m}^{(l,c)}
                \Biggl(
                    \left[
                        \mathbf{v}_{x,j}^{(l_1,c)}
                        \otimes
                        \mathbf{y}_{ji}^{(l_2)}
                    \right]_{m}^{(l)}
                    +
                    \left[
                        \mathbf{x}_j^{(l_1,c)}
                        \otimes
                        \mathbf{v}_{y,ji}^{(l_2)}
                    \right]_{m}^{(l)}
                \Biggr)
            \Biggr)
            \end{aligned}$

            \State $\displaystyle
            \begin{aligned}
            \mathbf{t}_{g,i}^{(l,c)}
            \mathrel{+}={}&
            w_{ji}^{(l_1,l_2,l,c)}
            \Biggl(
                \left[
                    \mathbf{q}_{x,j}^{(l_1,c)}
                    \otimes
                    \mathbf{v}_{y,ji}^{(l_2)}
                \right]^{(l)}
                +
                \left[
                    \mathbf{v}_{x,j}^{(l_1,c)}
                    \otimes
                    \mathbf{q}_{y,ji}^{(l_2)}
                \right]^{(l)}
            \Biggr)
            \\
            &+
            v_{w,ji}^{(l_1,l_2,l,c)}
            \Biggl(
                \left[
                    \mathbf{q}_{x,j}^{(l_1,c)}
                    \otimes
                    \mathbf{y}_{ji}^{(l_2)}
                \right]^{(l)}
                +
                \left[
                    \mathbf{x}_j^{(l_1,c)}
                    \otimes
                    \mathbf{q}_{y,ji}^{(l_2)}
                \right]^{(l)}
            \Biggr)
            \\
            &+
            q_{w,ji}^{(l_1,l_2,l,c)}
            \Biggl(
                \left[
                    \mathbf{v}_{x,j}^{(l_1,c)}
                    \otimes
                    \mathbf{y}_{ji}^{(l_2)}
                \right]^{(l)}
                +
                \left[
                    \mathbf{x}_j^{(l_1,c)}
                    \otimes
                    \mathbf{v}_{y,ji}^{(l_2)}
                \right]^{(l)}
            \Biggr)
            \end{aligned}$

            \State $\displaystyle
            \begin{aligned}
            t_{v_x,j,m_1}^{(l_1,c)}
            \mathrel{+}={}&
            \sum_{m=-l}^{l}
            \sum_{m_2=-l_2}^{l_2}
            C^{(l,m)}_{(l_1,m_1)(l_2,m_2)}
            \Bigl[
                g_{i,m}^{(l,c)}
                \Bigl(
                    w_{ji}^{(l_1,l_2,l,c)}
                    q_{y,ji,m_2}^{(l_2)}
                    +
                    y_{ji,m_2}^{(l_2)}
                    q_{w,ji}^{(l_1,l_2,l,c)}
                \Bigr)
            \\
            &\qquad+
                q_{g,i,m}^{(l,c)}
                w_{ji}^{(l_1,l_2,l,c)}
                y_{ji,m_2}^{(l_2)}
            \Bigr]
            \end{aligned}$

            \State $\displaystyle
            \begin{aligned}
            t_{v_y,ji,m_2}^{(l_2)}
            \mathrel{+}={}&
            \sum_{m=-l}^{l}
            \sum_{m_1=-l_1}^{l_1}
            C^{(l,m)}_{(l_1,m_1)(l_2,m_2)}
            \Bigl[
                g_{i,m}^{(l,c)}
                \Bigl(
                    w_{ji}^{(l_1,l_2,l,c)}
                    q_{x,j,m_1}^{(l_1,c)}
                    +
                    x_{j,m_1}^{(l_1,c)}
                    q_{w,ji}^{(l_1,l_2,l,c)}
                \Bigr)
            \\
            &\qquad+
                q_{g,i,m}^{(l,c)}
                w_{ji}^{(l_1,l_2,l,c)}
                x_{j,m_1}^{(l_1,c)}
            \Bigr]
            \end{aligned}$

            \State $\displaystyle
            \begin{aligned}
            t_{v_w,ji}^{(l_1,l_2,l,c)}
            \mathrel{+}={}&
            \sum_{m=-l}^{l}
            \Biggl(
                g_{i,m}^{(l,c)}
                \Biggl(
                    \left[
                        \mathbf{q}_{x,j}^{(l_1,c)}
                        \otimes
                        \mathbf{y}_{ji}^{(l_2)}
                    \right]_{m}^{(l)}
                    +
                    \left[
                        \mathbf{x}_j^{(l_1,c)}
                        \otimes
                        \mathbf{q}_{y,ji}^{(l_2)}
                    \right]_{m}^{(l)}
                \Biggr)
            \\
            &\qquad+
                q_{g,i,m}^{(l,c)}
                \left[
                    \mathbf{x}_j^{(l_1,c)}
                    \otimes
                    \mathbf{y}_{ji}^{(l_2)}
                \right]_{m}^{(l)}
            \Biggr)
            \end{aligned}$
        \EndFor
    \EndFor
\EndFor
\end{algorithmic}
\end{algorithm}

\clearpage

\section{Programming Interface}\label{app:programming-interface}
LunaTP replaces the gather--TP--reduce sequence with a single convolution call, using the same e3nn irreps, instructions, and per-edge weights. It also supports precision selection and optional envelope smoothing of the messages \citep{gasteiger2022directionalmessagepassingmolecular, fu2025learning, zhou2026matris, liao2026equiformerv3scalingefficientexpressive}. The code below illustrates the change in usage.

\par\addvspace{8pt}
\noindent\begin{minipage}{\linewidth}
\definecolor{tpKeyword}{HTML}{0000FF}
\definecolor{tpControl}{HTML}{AF00DB}
\definecolor{tpFunction}{HTML}{795E26}
\definecolor{tpType}{HTML}{267F99}
\definecolor{tpVariable}{HTML}{001080}
\definecolor{tpString}{HTML}{A31515}
\definecolor{tpComment}{HTML}{008000}
\definecolor{tpArrow}{HTML}{1B637F}
\lstset{
    language=Python,
    basicstyle=\ttfamily\small\color{black!90},
    keywordstyle=\color{tpKeyword},
    morekeywords=[2]{for,if,in,return},
    keywordstyle=[2]\color{tpControl},
    emph=[1]{index_select,index_add_,new_zeros,initialize,create_graph},
    emphstyle=[1]\color{tpFunction},
    emph=[2]{Irreps,TensorProduct,MatTP,LunaTP},
    emphstyle=[2]\color{tpType},
    emph=[3]{x,x_edge,edge_attr,weight,src,dst,tp,conv,graph,out,messages,irreps_in,irreps_edge,irreps_out,instructions},
    emphstyle=[3]\color{tpVariable},
    stringstyle=\color{tpString},
    commentstyle=\color{tpComment},
    showstringspaces=false,
    columns=fullflexible, keepspaces=true,
    breaklines=true, tabsize=4,
    numbers=none, frame=none,
    aboveskip=0pt, belowskip=0pt,
    lineskip=0.5pt
}
\tcbset{tpinterface/.style={
    colback=black!2, colframe=black!18,
    colbacktitle=black!2, coltitle=black!80,
    fonttitle=\sffamily\small\bfseries,
    boxrule=0.4pt, titlerule=0pt,
    sharp corners, boxsep=0pt,
    left=10pt, right=10pt, top=6pt, bottom=10pt,
    toptitle=8pt, bottomtitle=4pt,
    before skip=0pt, after skip=0pt
}}
\begin{tcolorbox}[tpinterface,title={e3nn}]
\begin{lstlisting}
from e3nn import o3

# 1.Initialize
e3nn_tp = o3.TensorProduct(
    irreps_in, irreps_edge, irreps_out, instructions,
    internal_weights=False, shared_weights=False,
)

# 2.Forward
x_edge = node.index_select(0, src)              # Gather
messages = e3nn_tp(x_edge, edge_attr, weight)   # Tensor product
if use_envelope:
    messages = messages * envelope              # Apply envelope
out = node.new_zeros((node.shape[0], e3nn_tp.irreps_out.dim))
out.index_add_(0, dst, messages)                # Reduce
\end{lstlisting}
\end{tcolorbox}
\par\nopagebreak\vspace{5pt}
{\centering
\begin{tikzpicture}[x=1mm,y=1mm]
    \filldraw[
        fill=white,
        draw=tpArrow,
        line width=1.2pt,
        line join=round
    ]
        (-1.7,6) -- (1.7,6) -- (1.7,2.8) -- (3.5,2.8)
        -- (0,0) -- (-3.5,2.8) -- (-1.7,2.8) -- cycle;
\end{tikzpicture}\par}
\nopagebreak\vspace{5pt}
\begin{tcolorbox}[tpinterface,title={LunaTP}]
\begin{lstlisting}
import LunaTP

# 1.Initialize from an e3nn TP or custom instructions.
lunatp_conv = LunaTP.initialize(e3nn_tp)

# 2.Forward
out = lunatp_conv(
    node, 
    edge_attr, 
    weight, 
    env=envelope, # Default: 'none' (no envelope)
    dtype='fp32'
)
\end{lstlisting}
\end{tcolorbox}
\end{minipage}
\par\addvspace{8pt}

Beyond the convolution formulation considered in this work, LunaTP supports a broader range of tensor products and their convolutional forms, including those using the \texttt{uuu}, \texttt{uvu} and \texttt{uvw} connection modes in e3nn.

\end{document}

%% file: math_commands.tex
\usepackage{amsmath,amsfonts,bm}

\def\eqref#1{equation~\ref{#1}}
\def\1{\bm{1}}

\DeclareMathAlphabet{\mathsfit}{\encodingdefault}{\sfdefault}{m}{sl}
\SetMathAlphabet{\mathsfit}{bold}{\encodingdefault}{\sfdefault}{bx}{n}